\documentclass[preprint, 12pt,  3p]{elsarticle}
\usepackage{setspace}
\usepackage{tcolorbox}
\usepackage{tikz}
\usepackage{array}
\usepackage{wrapfig}
\usepackage{amsmath, amsthm, amsfonts, cases}
\usepackage{graphicx}
\usepackage{mathtools}
\usepackage{adjustbox}
\usepackage{url}
\usepackage{array}
\usepackage{multirow}
\usepackage{caption}
\usepackage{subcaption}
\usepackage{color,soul}
\usepackage{setspace}
\usepackage{mathtools}
\usepackage{float}
\usepackage{adjustbox}
\usepackage{blkarray, bigstrut}
\usepackage{appendix}
\usepackage{enumerate}
\usepackage{hyperref}
\usepackage{ulem}
\usepackage[shortlabels]{enumitem}
\usepackage[left]{lineno}
\usepackage[linesnumbered,ruled,vlined]{algorithm2e}
\newtheorem{Ex}{Example}

\newcommand{\new}[1]{{#1}}
\pdfoutput=1
\begin{document}

\begin{frontmatter}
\title{Personalized
Summarization of Scientific Scholarly Texts}

\author[mainaddress]{Alka Khurana}
\ead{akhurana@cs.du.ac.in}
\address[mainaddress]{Department of Computer Science, University of Delhi, New Delhi 110007, India}

\author[mainaddress]{Vasudha Bhatnagar\corref{mycorrespondingauthor}}
\cortext[mycorrespondingauthor]{Corresponding author}
\ead{vbhatnagar@cs.du.ac.in}

\author[mainaddress]{Vikas Kumar}
\ead{vikas@cs.du.ac.in}

\begin{abstract} 

\textit{Personalized Summarization} aims to condense the given text such that it aligns with the user knowledge preferences while preserving the essential concepts and ideas present in the original document. In this paper, we present an unsupervised algorithm that creates an extractive summary of the given scientific scholarly text to meet the \textit{personal} knowledge needs of the user.  The proposed algorithm, \textit{P-Summ},  permits the user to supply both \textit{positive} and \textit{negative} knowledge signals and creates the document summary that ensconces content semantically similar to the positive signals and dissimilar to the negative signals. \new{The proposed approach employs Weighted Non-negative Matrix Factorization to unveil the latent semantic representations of the terms and sentences within a document. To incorporate user knowledge signals, we design a principled mechanism to set the weight matrix for factorization. This allows us to reward or penalize terms in the latent semantic space based on user-specified signals. Finally, each sentence is scored based on the contributions of the terms contained in it, and top-ranking sentences are selected for the\textit{ personal} summary.} 

We also propose a multi-granular evaluation framework to assess the quality of the  \textit{personal summary}   at three levels of granularity - sentence, term and semantic. The framework uses a high-quality generic extractive summary as a gold standard for evaluation to assess the extent of \textit{personalized} knowledge injected in the summary, thereby avoiding the potential human subjectivity in manually written gold standard summaries. \new{Using this framework, we conduct a comprehensive evaluation of \textit{P-Summ} algorithm on four publicly available datasets consisting of scientific scholarly articles. Our evaluation focuses on the impact of summary length and signal strength on the extent of \textit{personalization} of the summary. We report the results for summaries of varying lengths ($10\%$ - $25\%$ of the original document) and signal strengths ($1$ - $5$ keywords)}.

\end{abstract}


\end{frontmatter}

\section{Introduction}
\label{sec:introduction}
Significant progress has been observed in the field of automatic text summarization to generate \textit{generic} summaries of documents using unsupervised, supervised and semi-supervised methods. However, a generic summary may not always serve the purpose of augmenting users' existing knowledge due to differences in cognitive skills, background knowledge, and time-varying interests. This motivates research and development of algorithms, which generate \textit{personal} summary to meet the \textit{personal} knowledge needs of the user.  Recent shift towards designing algorithms to generate summaries that meet user's knowledge requirements demonstrates the increasing demand for algorithmically generated \textit{personal} summaries \citep{joint_optimization_avinesh2017, avinesh2018sherlock, ghodratnama2021adaptive, wikigaze_setia2021, dou2021gsum,  maddela2022entsum, he2022ctrlsum}. Resultingly, \textit{personalized} summarization is the next frontier in the field of automatic text summarization.

\textit{Personalized} summarization is the process of condensing the text  such that it accommodates the users' preferences for the content while maintaining the primary attributes of relevance and non-redundancy in the personal summary.   The task of  \textit{personalized} summarization entails two sub-tasks, viz., (i) modeling the user's knowledge preferences to be communicated to the algorithm and  (ii) creating a desired summary using either extractive or abstractive (neural) algorithm to best match the communicated knowledge preferences. 

Users' preferences for generating personal summaries are communicated to the algorithm in several different modes, including keywords, keyphrases, concepts, entities, queries, aspects, feedback, etc., which stand-in as\textit{ semantic signals}\footnote{In this paper, the phrases ``knowledge signals", ``knowledge needs", ``semantic signals" and ``user preferences" are used interchangeably.} to the summarization algorithm for inclusion or exclusion of content.   The diversity in modes for conveying user preferences to the system has driven personalized summarization research across a wide spectrum of approaches, including query-focused summarization, interactive or feedback-based summarization, aspect-based summarization, concept-based summarization, etc.

Query-focused summarization approaches accept user preferences in the form of query and extract the relevant content from the document to generate personal summary \citep{diaz2007user, park2008_personalized1, park2008_personalized2, query-focused-2009-zhao, query-focused-2013-luo, query-focused-2021-lamsiyah}. Interactive approaches accept user knowledge requirements in the form of clicks or feedback through an interface, and extract compatible content from the document for personal summary \citep{yan2011summarize, avinesh2018sherlock, ghodratnama2021adaptive, bohn2021hone}. The approach is intuitive and imparts freedom to the user to \textit{choose} the desired content. Aspect-based approaches generate a personal summary by focusing on aspects of the document specified by the user \citep{berkovsky2008aspect, frermann2019inducing, weak-supervision2020summarizing, soleimani2022zero-shot}. Aspect-based summarization is useful for quickly understanding the key findings in a document based on the perspective desired by the user. Concept-based summarization approaches consider the document as a set of concepts and extract the sentences that connote the concepts preferred by the user   \citep{boudin2015concept, ghodratnama2021adaptive}. Concept-based summarization comes in handy for understanding the theme of an article. Update summarization techniques result in a personal summary that is incremental and specifically condenses the new information that has been added to a document  \citep{delort2012dualsum-update, li2012update}.
 
Personalized summarization has been researched for a variety of genres including news articles \citep{ghodratnama2021adaptive, dou2021gsum, zhong2022unsupervised_granularity, he2022ctrlsum}, scientific articles \citep{dou2021gsum, he2022ctrlsum}, Wikipedia articles \citep{wikigaze_setia2021, dou2021gsum, hayashi2021wikiasp}, tweets \citep{tweet_chin2017totem}, reviews \citep{review_li2019towards}, opinions \citep{mukherjee2020read}, etc. The field has also been investigated for personalized summarization of a single document \citep{jones2002interactive, park2008_personalized1, park2008_personalized2, berkovsky2008aspect, yan2011summarize, azar2015query, frermann2019inducing, weak-supervision2020summarizing, dou2021gsum, he2022ctrlsum} and multiple documents \citep{boudin2015concept, avinesh2018sherlock, xu2020coarse, query-focused-xu2020generating, ghodratnama2021adaptive, query-focused-2021-lamsiyah, zhong2022unsupervised_granularity}. We give a detailed account of the approaches in Sec. \ref{sec:related-work}.

\new{\subsection{Personalized Summarization:  Challenges}}
\label{subsec:research-gap}
\new{The primary objective of \textit{personalized} summarization algorithm is to identify important information in conformity with the knowledge signals, that is, knowledge preferences specified by the user. Multifaceted growth in the area of personalized summarization has led to several modes of specifying the knowledge needs. This diversity in the modes for conveying knowledge preferences has sparked the following three major challenges.}

\begin{enumerate}[i.]
\item \new{{Eliminating Undesired Information from Summary: } The first challenge is to design a personalized summarization algorithm that is capable of eliminating undesired information from the summary. In the real world, individuals are often aware of the background knowledge possessed by them and mindful of the information \textit{required} or \textit{not required} in the summary. Thus, in order to personalize the summary, user may supply either  \textit{positive} or \textit{negative} knowledge signal to the algorithm. Positive knowledge signals connote the information \textit{desired} in the summary, while negative signals connote the information \textit{not desired}. }
 
\new{Existing algorithms for personalized summarization majorly consider only positive signals and extract semantically matching knowledge from the target document. These methods, typically neural,  are trained using triples of the form $<D,\mbox{ } \mathcal K,\mbox{ } S>$, where $D$ is the document to be summarized, $\mathcal K$ betokens knowledge desired by the user and $S$ is the gold standard reference summary of the document. \new{The inability to train models to handle negative knowledge signals is expected due to the impracticality of including negative training samples,  where the reference summary $S$ does not contain the negative knowledge signal. A subset of salient sentences that meets the summary length requirement and is semantically unrelated to the negative knowledge signals can be considered a valid summary. However, there can be multiple such subsets of sentences for a combination of keywords connoting negative knowledge signals, and  annotating subsets to select the best training sample is a challenging task. Therefore, constructing  training examples for   different combinations of negative knowledge signals is a humongous exercise. Under the circumstances, a neural model cannot learn to selectively ignore document content to generate a \textit{personal} summary that excludes the knowledge \textit{not desired} by the user.
 }}
 
\item \new{{Comparative Evaluation of Competing Algorithms: \label{challenge-ii}}
The second challenge arises due to the diversity in representational forms of user knowledge signals and manifests during comparative evaluation of personal summarization algorithms.  Different  \textit{modes} of semantic signals accepted  as user knowledge preferences obligate several different benchmark datasets (e.g. \cite{hayashi2021wikiasp, ahuja2022aspectnews, maddela2022entsum, yang2022oasum}).  Tight binding of the representational forms (modes)  of user knowledge preferences with the dataset limits empirical comparison of a new algorithm that accepts knowledge preference in a different mode. For example, comparison of summaries generated by an entity-based summarization algorithm with those generated by an aspect-based summarization algorithm  is incongruous due to inherent differences in the notion of \textit{entity} and \textit{aspect}. Similarly, an algorithm  performing aspect-based summarization cannot be evaluated using dataset curated for entity-based personalization algorithm.} 

\item \new{Evaluation Protocol:
The third challenge relates to the protocol employed for evaluating personal summaries. Since personal summaries are inherently subjective, their quantitative evaluation by humans is inefficacious. Traditional protocol for evaluating  algorithmic summaries against \textit{generic} gold standard reference summaries using ROUGE \citep{lin2004rouge} metric is not suitable for evaluating personal summaries because of the infeasibility of creating \textit{personal} gold summaries for multiple  knowledge signals. Semantic signals (user preferences) passed in different modes (e.g. keywords, keyphrases, aspects, concepts, queries) to a personalized summarization algorithm result in different personal summaries, necessitating creation of multiple reference summaries for evaluation. Clearly, this is clearly an insurmountable task.  }

\new{The popular protocol followed in several recent researches calls for extraction of user knowledge signals (keywords, keyphrases, aspects, etc.)  from the \textit{generic} gold standard reference summary and use them as positive knowledge signals for generating a \textit{personal} summary \citep{narayan2021planning, dou2021gsum, he2022ctrlsum}. This summary is expected to be richer in input knowledge signals than the system-generated \textit{generic} summary. Accordingly, a higher ROUGE score is expected compared to the latter, which is taken as evidence of the extent of  \textit{personalization}. The flip side of this protocol is that it promotes the use of \textit{only}  positive semantic signals for personal summary, completely failing to accommodate negative knowledge signals. }
\end{enumerate}

\new{\subsection{Contributions}}
\label{subsec:contributions}
\new{To overcome the above-mentioned challenges, we propose  \textit{P-Summ} algorithm that accepts keywords or phrases as \textit{positive} or \textit{negative}  knowledge signals to create a personal summary.  We also propose a multi-granular evaluation framework and demonstrate the effectiveness of \textit{P-Summ} algorithm for personal summarization of scientific scholarly articles, which are often lengthy and contain multiple points of interest. Our specific contributions are as follows.}
\begin{enumerate}[i.]
   \item \new{We propose a \textit{P}ersonalized \textit{Summ}arization algorithm, \textit{P-Summ}, which is competent to handle both \textit{negative} and  \textit{positive} knowledge signals specified by the user to create a personal summary. The algorithm elicits knowledge that is semantically related to the user-specified signals in the local and global context of the document and leverages it to meet user expectations in the personal summary (Sec. \ref{sec:proposed-algorithm}).  It can be adapted to handle other modes of knowledge signals like entities, aspects, queries, etc., by mapping them to the terms present in the document.} 
    \item  \new{We use Weighted Non-negative Matrix Factorization (wNMF) to expose the latent semantic space of the document and process both positive and negative knowledge signals in this space (Sec. \ref{sec:personalized-summarization}). To the best of our knowledge, this is the first attempt to handle negative knowledge signals to shove out undesired information from the summary.}
    \item \new{We propose a three-stage evaluation framework for assessing the quality of system-generated personal summaries, and use it to evaluate the effectiveness of \textit{P-Summ} algorithm (Sec. \ref{sec:evaluation-framework})}.
    \item \new{We state the experimental design in Sec. \ref{sec:experimental-design}, and present the results of a comprehensive evaluation of \textit{P-Summ} algorithm on four public datasets comprising scientific scholarly articles in Sec. \ref{sec:performance-evaluation-psumm}. We use the proposed  evaluation framework for the same and observe that the   \textit{P-Summ}  summaries meet user expectations.} 
\end{enumerate}

\begin{figure}[ht]
\centering
\includegraphics[width=0.95\textwidth]
{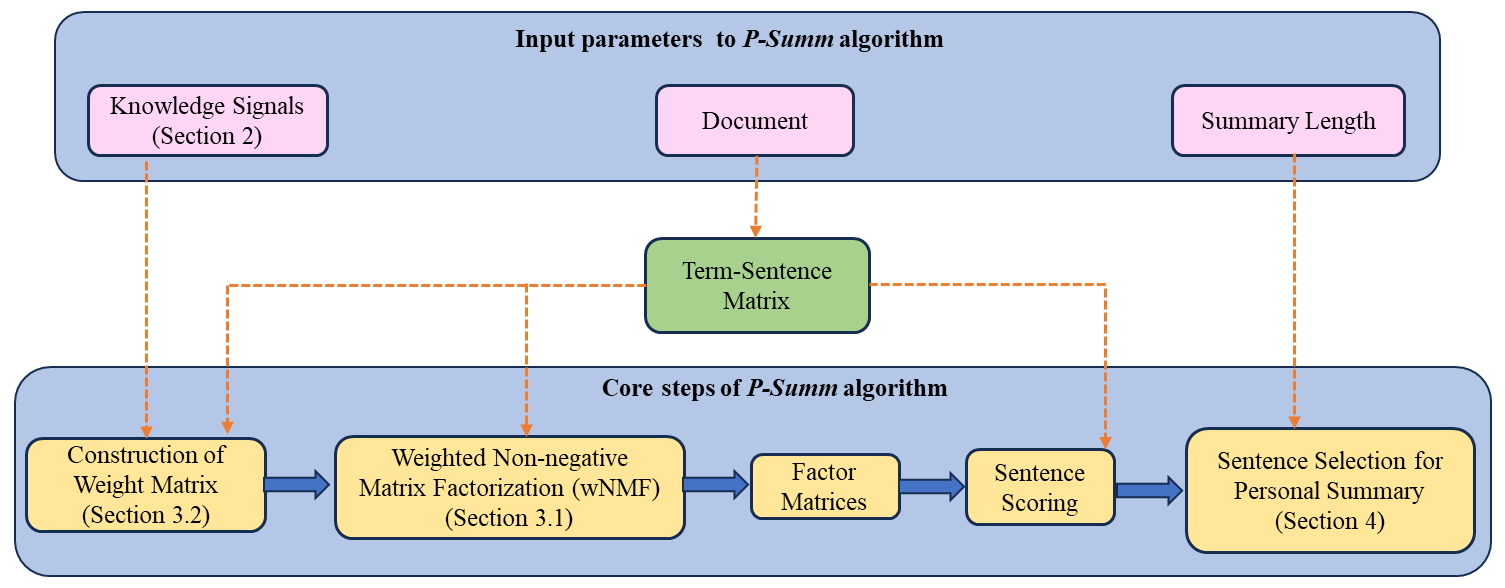}
\caption{Inputs to \textit{P-Summ} algorithm and its core steps}
\label{fig-psumm-pipeline}
\end{figure}

\new{Figure \ref{fig-psumm-pipeline} shows the inputs to the \textit{P-Summ} algorithm and the core processing steps. The user knowledge signals and binary term-sentence matrix of the document are used to construct the weight matrix, which plays an instrumental role in identifying information \textit{desired} and \textit{not-desired}  in the summary. Subsequently, the term-sentence matrix is factorized to reveal the latent semantic space of the document using Weighted Non-negative Matrix Factorization. The sentences are scored, and top-ranking sentences are selected to construct the summary of the specified length.}

\section{Personalized Summarization: The Problem and Approaches}
The aim of \textit{personalized} summarization is to automatically generate \textit{personal} summary of the document, which reflects the user's interests. The user communicates semantic signals ($\mathcal K$) pertinent to her topical interests for automatic generation of personal summary.  Let $\mathcal K^+$ denote positive knowledge signal, signifying the information that the user \textit{desires} in summary,  and  $\mathcal K^-$ be the negative knowledge signal to express the information  \textit{not required} in the \textit{personal} summary. Please note that we use    $\mathcal K$ to denote the knowledge signal in general, i.e., without preference or predisposition.

The problem addressed in the paper is formally stated as follows. Given the document $D$  and knowledge signal $\mathcal K$ ($\mathcal K^-$ or $\mathcal K^+$), the \textit{personalized} summarization algorithm generates the personal summary, $S_p$ of $D$, by congregating content corresponding to $\mathcal K^+$ and forsaking that corresponding to $\mathcal K^-$ from  $D$.  Ideally, the personal summary ($S_p$) is rich in $\mathcal K^+$  and devoid of  $\mathcal K^-$,  compared to generic summary ($S_g$), which prioritizes the content that is important in the global context of the document. Simply stated,  the personal summary is expected to contain diluted negative knowledge content and intense positive knowledge. 
\subsection{Related Works}
\label{sec:related-work}
Personalized summarization has been researched for single and multiple document settings, employing supervised and unsupervised techniques.  Algorithmic personal summaries can be customized by controlling parameters such as keywords, queries, summary length, high-level topics, entities mentioned in the document, and question-answers \citep{narayan2021planning, dou2021gsum, he2022ctrlsum, ahuja2022aspectnews}. More recently, the granularity of semantic coverage of the information articulated in the document has been explored by \citet{zhong2022unsupervised_granularity, 2023cocoscisum}. We categorize the approaches for personalized summarization on the basis of modes of user knowledge signals and types of summarization algorithms adopted to generate a personal summary of the document and describe them in detail below.

\subsubsection{Based on mode of Knowledge Signals}
\label{subsec:mode_of_knw_signals}
Various representational modes for expressing knowledge preferences include keywords, keyphrases, entities, aspects, concepts, queries, feedback, etc., which lead to the categorization of personalized summarization algorithms on the basis of the mode of knowledge signals.  Here, we outline the broad categories under which the problem has been researched.

\begin{enumerate}[i.]
\item \textit{Interactive Summarization}\\
These approaches facilitate the personalization of a document summary through the user interface via clicks or feedback. The user is presented with the preliminary summary of the document and then prompted to provide feedback. The method considers user feedback to extract the desired content and generate a personal summary of the document. The process continues iteratively until a personal summary that meets user knowledge preferences is generated. 

The interactive summarization approach proposed by \citet{jones2002interactive} allows the user to control summary length and topics of interest in the personal summary. The method extracts keyphrases from the document and assigns scores to them. The score of a sentence is computed by adding up the scores of keyphrases occurring in it. The user interacts with the system by choosing keyphrases to be included or excluded from the summary and assigning weights to them. The system re-scores the sentences according to the weights assigned by the user and selects top-scoring sentences for inclusion in the personal summary. \citet{yan2011summarize} propose an optimization framework that balances user interests and traditional summary attributes (coverage and diversity) to meet user expectations. \citet{avinesh2018sherlock} propose a \new {framework based on Integer Linear Programming for interactive multi-document summarization. The objective is to optimize concept coverage in the personal summary based on user needs.} The generated personal summary and the concepts contained in it are presented to the user iteratively, where the user is prompted to either accept or reject the concepts to further enhance the summary. \cite{ghodratnama2021adaptive} present a similar approach, which aims to maximize user-desired content selection. The system iteratively learns from users' feedback, where the user can accept or reject a concept (combination of words) included in the summary. Additionally, the user can change the importance of the concept and even decide the confidence level of her feedback.  \citet{bohn2021hone} propose a method that models user interests as a weighted set of concepts and maximizes the importance of a sentence such that selected sentences reflect the maximum weighted similarity to the user interests. 

\item \textit{Query-focused Summarization}\\
Query-focused Summarization (QFS) is one of the earliest summarization methods for creation of the personal summaries. QFS methods generate a personal summary of a document based on user knowledge signals communicated as natural language queries. The query is generic and may not be specific to any object, entity, or topic in the document. These systems require the user to have prior knowledge in order to frame the input inquiry.

Works in the area of QFS use queries as positive knowledge signals, that is, the information desired by the user in personal summary \citep{diaz2007user, park2008_personalized1, park2008_personalized2}. These works adopt the strategy of dual scoring of a sentence, where a sentence is scored for relevance with respect to the document and the user query. Subsequently, the two scores are combined using a weighted function to obtain the final scores for sentences, and top-scoring sentences are included in the personal summary. 
	
Query-focused Summarization has been majorly studied for multi-document summarization. \citet{query-focused-2009-zhao} propose a multi-document graph-based QFS algorithm that uses sentence-to-sentence and sentence-to-word relations to expand user query and then employ a graph-ranking algorithm to select the highest scoring sentence from the document that answers the query. \citet{query-focused-2013-luo} employ a probabilistic framework to model relevance, coverage, and novelty for query-focused multi-document summarization. Subsequently, a greedy sentence selection method is employed for selecting sentences for summary. Recent work by \citet{query-focused-2021-lamsiyah} leverages pre-trained sentence embedding models and propose a method for query-focused multi-document summarization. Sentence embedding models are used for computing dense vector representation of sentences and user queries, and similarity between two values is computed. Subsequently, top-scoring sentences are selected from the document based on their relevance to the query.

\item \textit{Aspect-based Summarization}\\
Aspect-based Summarization is the task of summarizing a document for a specific point of interest. An early work by \citet{berkovsky2008aspect} partitions the document based on the aspects and utilizes a 4-point scale of interest for aspects. Subsequently, the content for a given aspect proportional to the user's interest in the aspect is selected for inclusion in the summary. In the last few years, aspect-based summarization has been researched using neural approaches described in Sec. \ref{subsubsec-based-on-summarization-approach}.
\end{enumerate}	

\subsubsection{Based on Approach for Summarization}
\label{subsubsec-based-on-summarization-approach}
In recent years, personalized summarization has been investigated using neural approaches. These approaches train summarization models using the document to be summarized, knowledge signals, and gold reference summary of the document. Subsequently, the trained neural model is employed to generate a personal summary of the document to accommodate the user's knowledge preferences.

\citet{azar2015query} present a query-based single document summarization method that learns a condensed representation of the document using an ensemble of noisy auto-encoders, which in turn leverages the query to select relevant parts of the document for the final summary. \citet{frermann2019inducing} follow a neural encoder-decoder architecture with attention for the task of aspect-specific summarization, which is capable of generating abstractive and extractive aspect-driven summaries of the document. \citet{weak-supervision2020summarizing} propose a neural model for abstractive aspect-based summarization, which leverages external knowledge sources such as WordNet and Wikipedia for enriching weak supervision in training and creates a robust model to summarize any aspect. 

\begin{figure}[ht]
\centering
\includegraphics[width=0.95\textwidth]{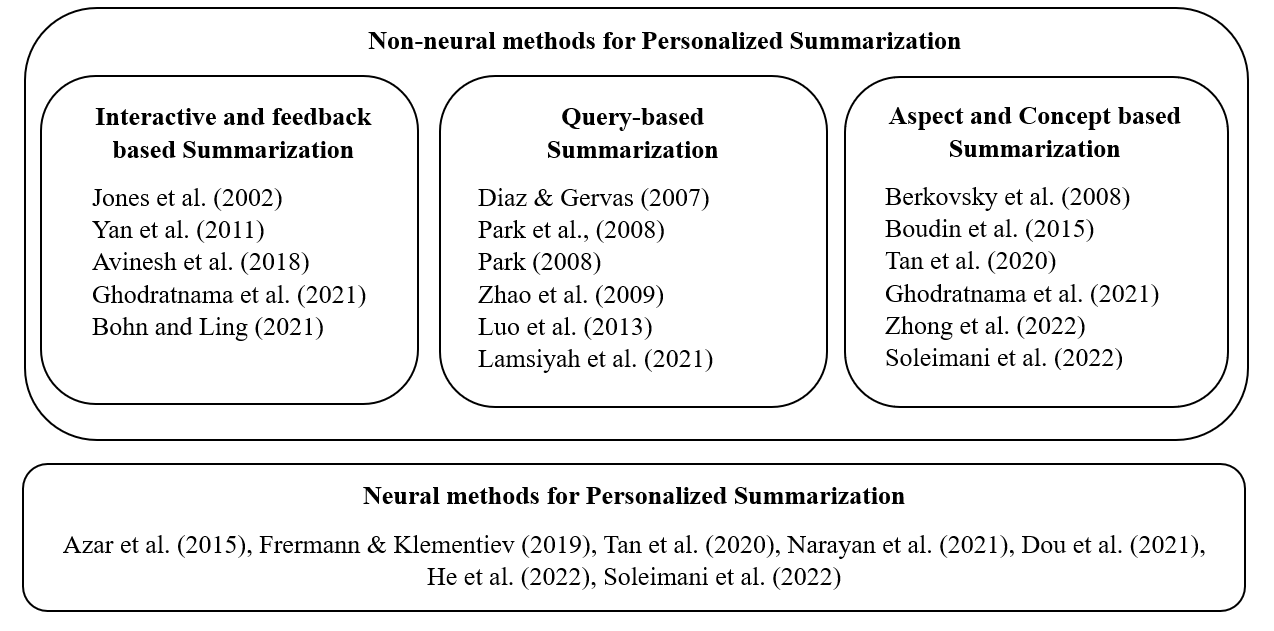}
\caption{Approaches for Personalized Summarization}
\label{fig-approaches-personal-summ}
\end{figure}

The method proposed by \citet{narayan2021planning} employs a transformer-based sequence-to-sequence model for abstractive personalized summarization. The model extracts entity chains during training, which are ordered sequences of key entities from reference summaries. These entity chains serve as a prompt to guide the summary generation process, enabling the model to produce summaries that are more focused on essential entities and less likely to contain irrelevant information. This technique enhances the model's ability to summarize unseen documents effectively.
\citet{dou2021gsum} employ a neural-based transformer architecture, leveraging Oracle summaries to extract diverse forms of external guidance, such as keywords, relations, sentences, etc., for training the model. This guidance enhances the model's ability to generate coherent abstractive summaries. 
The method proposed by \citet{he2022ctrlsum} leverages neural-based BART architecture to generate controllable summaries of the text. The model is fed with input consisting of the document, generic reference summary, and keywords extracted from the summary during the training phase. The model learns to maximize the ROUGE score with respect to the generic reference summary. Subsequently, the trained model is employed to generate the personal summary of the text. The model can be adapted to other controlling parameters, such as entity, length, question-answering, etc., without requiring model retraining. \citet{soleimani2022zero-shot} propose a zero-shot model for aspect-based summarization and employ self-supervised pre-training to improve the performance on unseen aspects. Figure \ref{fig-approaches-personal-summ} consolidates the approaches proposed for personalized summarization.
 
\textit{Existing approaches for personalized summarization are incapable of eliminating knowledge that is not interesting to the user. To the best of our knowledge, \textit{P-Summ} is the first algorithm that can accommodate both positive and negative knowledge preferences of the user. Unlike most existing  algorithms, \new{P-Summ is adaptable to handle other modes of knowledge signals, including entities, aspects, and queries}. }

\section{Personalized Summarization - Handling Positive and Negative signals}
\label{sec:personalized-summarization}
A \textit{personal} summary of a document aims to capture the vital content from the document to meet the topical knowledge needs of the user, which may be expressed as knowledge signals ($\mathcal K$). Existing background knowledge, which is banal for the user and \textit{not desired} in the summary, is explicitly expressed as $\mathcal K^{-}$, whereas the new knowledge \textit{desired} in the personal summary is represented as $\mathcal K^{+}$. The proposed personalized summarization algorithm down-scores the sentences that semantically match $\mathcal K^{-}$ and up-scores those matching $\mathcal K^{+}$. 

Down-scoring sentences that convey background knowledge of the user can be effectuated by penalizing  $\mathcal K^{-}$ either directly in the document space or in the latent semantic space of the document. Since a sentence in the document space may possibly convey more than one concept, simply eliminating sentences containing negative signals is an extreme penal action.  Thus, a strategy to directly penalize a sentence in the document space is naive because it inadvertently blocks other important information conveyed in the sentence. Ergo, inspecting the latent semantic space is a compelling need for bypassing knowledge \textit{not desired} by the user. Similarly, scouring the document space and simply choosing sentences containing  $\mathcal K^{+}$ is also a naive action, since it may result in the inadvertent inclusion of redundant information. Further, it is prudent to identify sentences that not only convey desired knowledge ($\mathcal K^{+}$), but also semantically related information. Therefore,   analysis of the latent semantic space of the document also benefits up-scoring of sentences that comprise positive knowledge signals.

Conventional techniques to uncover latent space of the document include Latent Semantic Analysis (LSA) \citep{LSA1990deerwester}, Nonnegative Matrix Factorization  (NMF) \citep{NMF1999nature}, Latent Dirichlet Allocation (LDA) \citep{LDA2003latent} etc. We choose to go with the relatively recent method of Weighted Non-negative Matrix Factorization \citep{WNMF2009kim} because of its ability to divulge the importance of terms and sentences in the latent semantic space of the document in accordance with the weight matrix. Weighted NMF empowers the proposed \textit{P-Summ} algorithm to penalize (or reward) the keywords in $\mathcal K^{-}$ (or $\mathcal K^{+}$) individually, based on how much they contribute to the latent topics in the document.

\subsection{Weighted Non-negative Matrix Factorization}
\label{sec:weighted-nmf}	
Consider the term-sentence matrix $A_{m \times n}$ representing document $D$, containing $n$ sentences $\{S_1, S_2, \dots, S_n\}$ and $m$ terms $\{t_1, t_2, \dots, t_m\}$. Rows in matrix $A$ correspond to terms in $D$, columns represent the sentences, and element $a_{ij} \in A$ represents the occurrence of term $t_i$ in sentence $S_j$.  Let $W_{m \times n}$ be a matrix of non-negative weights, where the entry $w_{ij}$ controls the prominence of the element $a_{ij} (\in A$) in the latent space.  Weighted Non-negative Matrix Factorization (wNMF) decomposes the matrix  $A_{m \times n}$ into two non-negative factor matrices $U_{m \times r}$ and $V_{r \times n}$ such that the objective function, ${\mid\mid W\odot(A- UV) \mid\mid}^2_F$ is minimized. Here, $\odot$ denotes Hadamard product, and $r (\ll m, n)$ is the number of latent factors ($\tau_1, \tau_2, \dots, \tau_r$) into which the document $D$ is decomposed. Matrix $W$, when initialized with all 1's, leads to standard non-negative matrix factorization.

In the current context, matrix $A$  is the binary incidence matrix of the document, matrix $U$ captures the relationship among the terms and latent factors, and matrix $V$ represents the relationship among the latent factors and sentences in the latent semantic space of $D$. Element $u_{ij}$ in $U$ denotes the strength of term $t_i$ in the latent topic $\tau_j$, and element $v_{ij}$ in  $V$ represents the strength of latent topic $\tau_i$ in sentence $S_j$. These representations maintain conformity with the weights assigned by matrix W to the terms in $A$.

\subsection{Penalty and Reward for Knowledge Signals}
\label{sec:penalty-reward}
Recall that the generation of personal summary necessitates regulating the sentence score to adapt summary to the user-specified knowledge signals ($\mathcal K^- \mbox{or } \mathcal K^+$). \textit{Penalizing} and \textit{rewarding} the constituent terms of sentences via wNMF is a viable strategy to control sentence scores appropriately. The weights in matrix $W$ are exploited to penalize terms in $\mathcal K^-$ and
reward those in $\mathcal K^+$ in the latent semantic space. The topic strength ($v_{ij}$) and term
strength ($u_{ij}$)  in the resulting factor matrices are effective indicators of the content preferred for exclusion or inclusion in the personal summary. Since weights in $W$ correspond to the terms occurring in $D$, we fragment keyphrases in the knowledge signals into uni-grams (terms).

\subsubsection{Identify Semantically related Terms}
\label{sec-identify-related-terms}
An effective personal summary must not only avoid (or include) the content directly conveyed by the user-specified knowledge signals, but must also consider other semantically related content in the document.   To affect this, we make use of the established premise that terms co-occurring in a sentence are semantically related to each other and tend to purport closely linked meanings \citep{gerlach2018network_co_occurrence}. Accordingly, we identify terms that co-occur with those comprising user knowledge signals.  Collectively, these terms are semantically related to the knowledge signals specified by the user and form a cohort.  They all must be penalized or rewarded concomitantly, albeit to the extent to which they contribute to the topics comprising the user knowledge signals. 

Semantic relatedness of terms in a sentence is elegantly captured by the graph representation of the text \citep{sonawane2014graph}. Accordingly, we create an undirected, weighted co-occurrence graph $\mathcal G = (N, E, F)$ of the terms in document $D$. Here, $N$ is the set of nodes and node $n_{i} \in N$ corresponds to the term $t_i \in D$. \new {$E (\subseteq N\times N)$ is the set of  edges  and $F$ is weighted adjacency matrix for $\mathcal G$. Element  $f_{ij}$ in $F$  is the weight of the edge $e_{ij}$ and denotes the frequency with which  terms $t_i$ and $t_j$  co-occur  in $D$.}

\begin{figure}[ht]
\centering
\adjustbox{max width=\linewidth}{
\includegraphics[width=\linewidth]{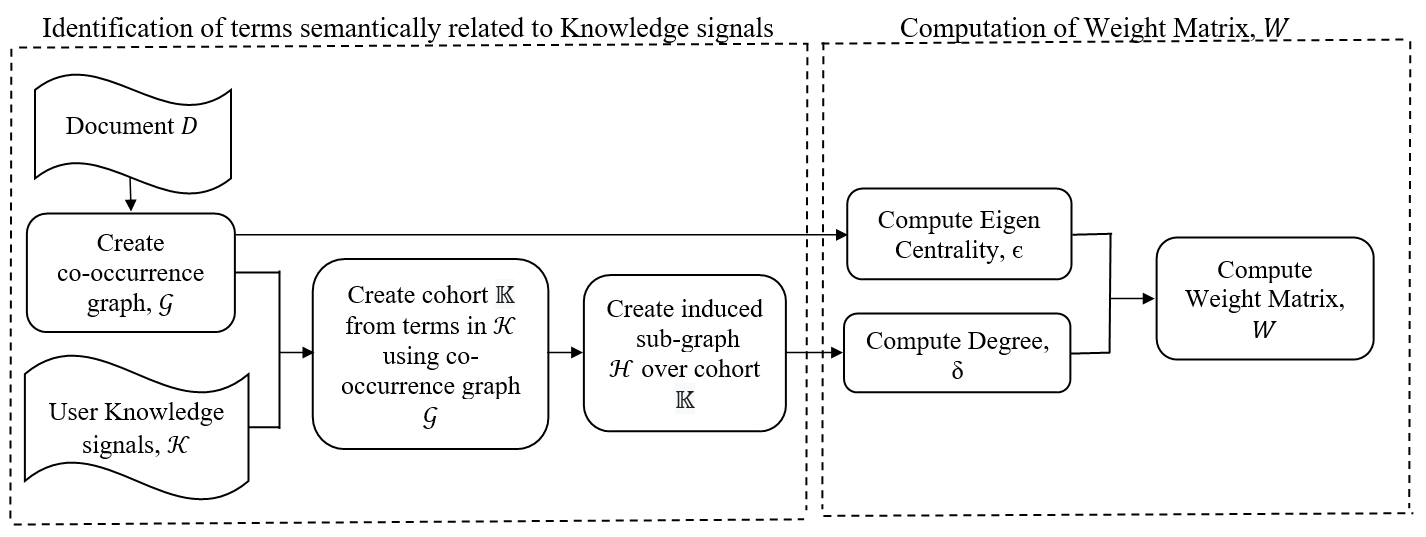}
}
\caption{Approach for computing weight matrix $W$}
\label{fig-Weight-function-pipeline}
\end{figure}

A schematic of the approach for identifying terms semantically related to user knowledge is shown in the left box in Fig. \ref{fig-Weight-function-pipeline}.  Given the document $D$ and user knowledge signal $\mathcal K$, we first create the co-occurrence graph  $\mathcal G$ of $D$. We then augment $\mathcal K$ by adding one-hop neighbours of all terms contained in it to yield the augmented set $\mathbb K$ denoted as $one\mbox{-}hop\mbox{-}neighbours(\mathcal K)$. Effectively, $\mathbb K$ is the cohort of all terms that are semantically related to the user knowledge signal. The sub-graph $\mathcal H$ induced from  $\mathcal G$  using terms in $\mathbb K$ betokens the semantic structure of user knowledge signal and related terms. The two graphs, that is, co-occurrence graph $\mathcal G$ and induced sub-graph $\mathcal H$,   are instrumental in accommodating user preferences for the personal summary in the latent space. In the figure,  induced sub-graph $\mathcal H$ symbolizes $\mathcal H^-$ (or $\mathcal H^+$),  corresponding to user knowledge signals $\mathcal K^-$ (or $\mathcal K^+$). In the interest of clarity, we summarize the notations introduced so far in Table \ref{table-notations}.

\begin{table}[ht]
        \centering
        \caption{Notations for processing of knowledge signals}
		\begin{tabular}{l|l}
        \hline
		\textbf{Symbol} & \textbf{Description}\\
			\hline
		      \centering{$\mathcal K$}  & Knowledge signals specified by the user (knowledge \textit{not desired} or \textit{desired} \\
                ($\mathcal K^-$ or $\mathcal K^+$)&in the summary)\\
                \hline
                \centering{$\mathcal G$}&Co-occurrence graph of the document $D$\\
			\hline
                \centering{$\mathbb K$} & Set of terms in $\mathcal K$ and semantically related terms in $D$ \\
                \hline
                $\mathcal H$ & \multirow{2}{*}{Sub-graph of $\mathcal G$, induced from terms in  $\mathbb K$ (corresponding to $\mathcal K^-$ or $\mathcal K^+$)}\\ ($\mathcal H^-$ or $\mathcal H^+$) & \\
                \hline
		\end{tabular}
		\label{table-notations}
\end{table}

\subsubsection{Setting the Weight Matrix}
\label{sec-weight-matrix}
The next step is to regulate the weight of terms in $\mathbb K$ for penalization (or reward) in the latent semantic space. We design a principled mechanism to set matrix  $W$, such that the weights for terms  (knowledge signals)  in $\mathcal K^-$ are less than one, and those in $\mathcal K^+$ are greater than one. Furthermore, weights must be in consonance with the importance of the terms in the \textit{global} and the \textit{local} contexts of the document, ensconced in  $\mathcal G$ and $\mathcal H$, respectively.
 
A term is important if it co-occurs with other important terms in the document.  \textit{Eigenvector centrality} of a node in the co-occurrence graph $\mathcal G$ encapsulates this idea \citep{zaki2014data}. In the current context, \textit{eigenvector centrality }of the nodes in $\mathcal G$ quantifies the importance of terms in the global context of the document $D$. Given the term $t_i \in \mathbb K$ and the corresponding node $n_i \in \mathcal G$, we denote its eigenvector centrality by  $\epsilon_i$ signifying the global importance of $t_i$.  $\mathcal H$ condenses the local context of the terms in the document, and it has a diameter of two due to its construction. Therefore,  computing \textit{eigenvector centrality} of the nodes is not meaningful.  \textit{Degree} $\delta_i$ of the  node (term $t_i$)  indicates the importance of node  in local context. Here, $\delta_i$  is computed using $\mathcal H$ ($\mathcal H^-$ or $\mathcal H^+$) based on the $t_i$ under consideration.  A node with a higher degree in $\mathcal H$ co-occurs more frequently with terms that are semantically close to it. Each term $t_i \in   \mathbb K$ is assigned weight, which is a function of $\delta_i$ (importance in the local context) and $\epsilon_i$ (importance in global context). The right box in Fig. \ref{fig-Weight-function-pipeline} outlines the approach for computing the weight matrix, $W$.
 
It is important to note that for any extractive personal summarization method,  penalization (or reward) of terms in $\mathbb K$ is instrumental in down-scoring (or up-scoring) the sentences containing these terms. For this reason, design of the weighting function is critical to the effectiveness of the proposed personal summarization algorithm. Conjecturing that the penalization (or reward) of the terms should be in conformity with their importance in both local and global context of the document text, we propose the following weighting scheme.

Consider the term $t_i \in \mathbb K$ occurring in sentence $S_j$. Since wNMF authorizes setting of the weight of terms in each sentence, we need to set $w_{ij} = 1$ if $t_i$ does not appear in $S_j$. Otherwise, the weight is set to less (or more) than one if the user has specified the term as a negative (or positive) knowledge signal. Equations \ref{eqn-penalty} and \ref{eqn-reward} quantify the weight of term $t_i$ in sentence $S_j$ for negative and positive knowledge signals, respectively. The computation accounts for both local and global importance of the term $t_i$ using its eigenvector centrality ($\epsilon_i$) in $\mathcal G$, and degree $\delta_i$ in $\mathcal H$.

\begin{subnumcases}{w_{ij} = }
    e^{-{\delta_i* \epsilon_i}*I(t_i \in S_j)},\mbox{ } \text{if $t_{i} \in one\mbox{-}hop\mbox{-}neighbours( \mathcal K^-$)} \label{eqn-penalty}\\
    e^{+{\delta_i* \epsilon_i}*I(t_i \in S_j)},\mbox{ }\text{if $t_{i} \in one\mbox{-}hop\mbox{-}neighbours( \mathcal K^+$)} \label{eqn-reward}
\end{subnumcases}

The use of $I(\cdot)$, the indicator function (1 if term $t_i \in S_j$, zero otherwise) guarantees that $w_{ij} = 1$ in case the term does not appear in the sentence.  Post-factorization, the strengths of terms and sentences in the latent semantic topics are reduced (or boosted) to favour their exclusion (or inclusion) in the personal summary based on the type of user knowledge signal. Long documents result in large graphs, which lead to large values of eigenvector centralities. In such cases,  the exponents in Eq. \ref{eqn-reward} for terms in positive knowledge signal result in exaggerated weights.  Normalization, in such situations is necessary to ensure that the weight values do not interfere with convergence of wNMF.

\begin{algorithm}
	\KwIn{Term-sentence matrix $A_{m \times n}$, Co-occurrence graph $\mathcal G$, User knowledge signal $\mathcal K$ ($\mathcal K^{-} \mbox{or } \mathcal K^{+}$)}
	\KwOut{Weight Matrix, $W_{m \times n}$ }
	$W$ $\leftarrow$ $[1]_{m \times n}$\;
	$\Vec{\epsilon}$ $\leftarrow$ Eigenvector-Centrality($\mathcal G$)\tcp*[l]{eigenvector centrality of the terms in $\mathcal G$}   
		$\mathbb K \leftarrow one\mbox{-}hop\mbox{-}neighbours(\mathcal K)$\;
		$\mathbb K \leftarrow \mathbb K \cup \mathcal K$\tcp*[l]{cohort of terms in user signals $\mathcal K$}
	    $\mathcal H$ $\leftarrow$ Induced-Subgraph($\mathbb K$) \tcp*[l]{induced subgraph over cohort $\mathbb K$}
		$\Vec{\delta}$ $\leftarrow$ Degree($\mathcal H$)\tcp*[l]{degree of the terms in $\mathcal H$}
		\ForEach{$t_{i} \in \mathbb K$}{
			\For{$j\gets1$ \KwTo $n$}
			 {
			     \If{$I(t_i \in S_j)$}{
					Compute $w_{ij}$ using Eq. \ref{eqn-penalty} to penalize 
                    (Eq. \ref{eqn-reward} to reward)
                    }}}	\caption{Algorithm for computing weight matrix, $W$}
\label{algo-weight}
\end{algorithm}
Algorithm \ref{algo-weight} presents the procedure for computing the weight matrix $W$. Input parameters to the algorithm are term-sentence matrix $A$ for the document to be summarized, its co-occurrence graph $\mathcal G$,   and knowledge signals $\mathcal K$ ($\mathcal K^- \mbox{ or } \mathcal K^+$).  Weight matrix $W$, which has the same dimensions as $A$, is initialized with all ones in step 1.  Step 2 computes the eigenvector centrality of nodes in graph $\mathcal G$. Step 3 retrieves one-hop neighbours for the terms in user signals $\mathcal K$  and denotes the resultant set by $\mathbb K$. Step 4 augments set $\mathbb K$ with the terms in knowledge signal $\mathcal K^-$ ($\mathcal K+$) given by the user and creates the cohort of terms representing the knowledge \textit{not desired} (or \textit{desired}) in summary. In step 5, the algorithm creates the induced sub-graph $\mathcal H$ for terms in cohort $\mathbb K$, and step 6 computes the degree of nodes. The loop in steps 7-10 computes the weights of the terms (penalty or reward) \textit{not desired} or \textit{desired} by the user in the personal summary.
\subsection{Number of latent factors}
\label{subsec:number-of-latent-factors}
The number of latent factors, $r$, into which the term-document matrix $A$ is to be decomposed determines the dimensions of the latent semantic space of $D$.  This is a decisive hyper-parameter for weighted non-negative matrix factorization of $A$. Ideally, the number of dimensions for factorization of $A$ corresponds to the number of core concepts (topics) described in the document.   Decomposing into more than the optimal number of topics diffuses focus of the topics, while decomposing into lesser topics leads to multiple concepts being addressed in a topic. Both situations can potentially degrade the sentence selection for the personal summary. Finding optimal number of latent topics in a text document is a challenging task because it depends on several factors, including writing style adopted by the author, desired summary length, and the background knowledge of the writer. 

Early methods for extractive summarization using NMF put the onus of deciding the number of latent factors on the user \citep{park2008_personalized1, park2008_personalized2, lee2009automatic}. \citet{alguliev2011mcmr} compute the number of latent topics as the ratio of number of unique terms to the total terms in the document. This method is straightforward and takes into account term-level information in the document. However, the method ignores the relationship between terms, which is pertinent for computing the number of topics. \citet{khurana2021investigating} employ community detection algorithm in the graph representation of text and use the number of detected communities as the number of latent topics (concepts). The method captures the semantic relatedness of the terms, which is relevant for identifying latent topics in the document. 

Since a topic in the document is articulated as the set of semantically related and frequently co-occurring terms, the idea of the number of detected communities over $\mathcal G$ serving as the number of latent topics in document $D$ is appealing. Following this approach, we set $r$ to the number of communities detected in  $\mathcal G$ using  Louvain algorithm \citep{Louvain2008blondel}, which is an efficient community detection algorithm and is parameter-free.   

\section{P-Summ – The Proposed Algorithm}
\label{sec:proposed-algorithm}
Post-factorization,  matrices $U$ and $V$  relate the \textit{terms-sentences-topics} in the latent semantic space of the document. The two matrices mask the penalties and rewards of the terms and can be effectively employed to select sentences for the personalized summary.  Sentences that are semantically close to the knowledge expressed in  $\mathcal K^{-}$ must be avoided for inclusion in the personal summary, while those which are close to $\mathcal K^{+}$  should be treated favourably. 

 Since the role of the terms corresponding to user knowledge signal for generating personal summary is paramount, we follow the term-based sentence scoring method proposed earlier by \cite{khurana2019extractive}. The method, NMF-TR,  is grounded on the principle that the terms with higher contribution to the latent topics of the document construe the content of the document better than those with lower contribution. NMF-TR  considers the sentence score as the sum of contributions of terms contained in it and efficiently computes the sentence score as follows. 
\begin{equation}
	\label{eqn-nmf-tr}
	Score(S_j) = \sum\limits_{i=1}^m  \phi_{i}*I(t_i\in S_j), 
\end{equation} 
\noindent
where $\phi_{i}$ represents normalized contribution of term $t_i$ in $r$ latent topics and is computed from the term-topic matrix $U$. If $u_{iq}$ ($\in U$) denotes the strength of term $t_i$ in latent topic $\tau_q$, then
\begin{equation}
	\label{eqn-phi}
\phi_i = \frac{\sum\limits_{q=1}^r u_{iq}} {\sum\limits_{p=1}^m\sum\limits_{q=1}^r u_{pq}}
\end{equation} 
The indicator  $I(.)$ in Eq. \ref{eqn-nmf-tr} denotes the occurrence of term $t_i$ in sentence $S_j$. 

\begin{algorithm}
	\KwIn{Document $D$, User knowledge signals $\mathcal K$ ($\mathcal K^- or \mathcal K^+$), Summary-length $\mathcal{L}$}
	\KwOut{Personal Summary, $S_p$ }
	Pre-process $D$ and create term-sentence matrix $A$\;
	Create co-occurrence graph $\mathcal G$ for document $D$\;
	Apply community detection to compute number of topics, $r$, in $D$\;
	Use $A$, $\mathcal G$, $\mathcal K$ to create weight matrix $W$ using Algorithm \ref{algo-weight}\; 
	$U$, $V$  $\leftarrow$ wNMF($A, W, r$)\; 
	Score the sentences in $D$ using Eq. \ref{eqn-nmf-tr}\;
	Select top-scoring sentences to generate personal summary, $S_p$ of length $\mathcal{L}$\;
	\caption{P-Summ Algorithm}
	\label{proposed-algo}
\end{algorithm}

Algorithm \ref{proposed-algo} outlines the proposed \textit{P-Summ} algorithm, which accepts three inputs, viz. document $D$, user's knowledge $\mathcal K$ and summary length $\mathcal{L}$. Step 1 pre-processes document $D$ by removing punctuation, stop-words and performing lemmatization\footnote{We employ spaCy's ``en\_core\_web\_sm'' model for lemmatization.}  to create term-sentence matrix $A$. Step 2 creates a co-occurrence graph $\mathcal G$ of the terms in $D$. In the next step, we compute the number of latent topics $r$ using Louvain algorithm \cite{Louvain2008blondel}. Step 4 sets the weight matrix $W$ using Algorithm \ref{algo-weight}. The next step decomposes   $A$, weighted by $W$, using wNMF with $r$ topics. In step 6, \textit{P-Summ} scores the sentences in $D$ using Eq. \ref{eqn-nmf-tr}. Finally, top-scoring sentences are selected for inclusion in the algorithmic personal summary while delimiting the summary length. 

\section{Evaluation Framework for Personal Summaries}
\label{sec:evaluation-framework}
Evaluation of a \textit{personal}  summary requires addressing two challenging issues. First, \textit{creating a gold standard personal summary,} and second, \textit{using effective metric to quantify the extent of personalized knowledge injected in the resulting personal summary}. The first one is referred as the evaluation protocol issue, and the second is simply the metric issue. The proposed framework relies on a high-quality system-generated \textit{generic} extractive summary ($S_g$) to resolve the protocol issue and employs three metrics to quantify the extent of new knowledge injected in the \textit{personal} summary ($S_p$).  We elaborate the two issues and our solutions in the rest of the section.
\subsection{Evaluation Protocol}
\new{We argue that the conventional protocol for evaluating algorithmic summary against the gold standard reference summary is infeasible for the evaluation of the personal summary.  The reason is that the user can specify any combination of keywords as positive and negative knowledge signals, and the conventional protocol demands the corresponding gold standard summary for evaluation. \new{Considering the document as the universal set of signals (ideas or concepts), those present in the generic gold standard can be considered as positive signals, and those which are not present as negative signals. Paraphrasing of the reference summary does not alter the set of contained positive signals and omitted negative signals,  and hence it cannot be used for evaluating a personal summary for a different set of signals. Practically, a gold standard reference summary is required for each possible combination of user signals. Curating reference summaries for all combinations of user knowledge signals is clearly infeasible. } }

\new{We believe that in the setting of \textit{personalized} summarization, comparison of the system-generated \textit{personal} summary ($S_p$) against system-generated \textit{generic} summary ($S_g$) is a prudent approach for quality assessment. This comparison objectively divulges the extent to which the \textit{desired} knowledge has been introduced or the  \textit{undesired} knowledge has been eliminated from the generic summary. This strategy completely insulates the quality evaluation of a personal summary from variations due to subjectivity (bias) in a human-written reference summary. Furthermore, since \textit{P-Summ} algorithm is an extractive algorithm, we favour extractive generic summary for evaluating a personal summary for all combinations of signals. With recent advances in the field of automatic extractive summarization   \citep{gambhir2017recent, survey_el2021automatic}, it is straightforward and economical to extract high-quality generic summaries.}
 
\subsection{Evaluation Metric} 
\label{subsec:eval-metric}
\new{To address the metric issue, we propose an evaluation strategy operating at three levels of granularity.  The first and the second stages quantify the lexical overlap between personal summary (system-generated) and generic gold standard summary at the sentence and term levels, respectively.  These two stages are indicative of injection of new knowledge in the personal summary in comparison to the generic summary. Scrutiny in the latent semantic space, the finest granularity level, confirms the extent of personalization of the summary. Note that sentence-level and term-level tests are preliminary tests for content change at coarse and meso levels, respectively.  However, neither test delves into the semantics of the content change. We delineate the three stages and the corresponding metrics below.} 
\begin{enumerate}[i.]
\item {Sentence-level Evaluation: }
\label{subsec:sentence-level-evaluation}
\new{This is the preliminary test to quantify the number of new sentences pushed in the summary compared to the generic summary and is a basic indicator of the performance of the algorithm.  Considering summaries $S_p$ and $S_g$ as sets of sentences, the extent of personalization achieved by the algorithm is indicated by the overlap of sentences between $S_p$ and $S_g$.  In case the user specifies $\mathcal K^-$, sentences semantically matching  $\mathcal K^-$ are displaced from $S_g$ in favor of other important ideas in the document. In the case of positive signals,  sentences related to  $\mathcal K^+$ are introduced in $S_p$ after nudging out those sentences from $S_g$ that are unrelated to  $\mathcal K^+$.  Therefore, the personal summary $S_p$ bears low sentence overlap with the generic summary $S_g$ for both types of signals, which indicates the introduction of new content in $S_p$. Since the construction of the weight matrix takes into account the local and global context of the user-given knowledge signals, we expect the new content to be in consonance with the given knowledge signals. } 

\new{We use Jaccard Index (JI) to gauge the quantum of new knowledge introduced in the personal summary relative to the generic one. A common metric to quantify overlapping between two sets, here JI is an indicator of incorporation of user knowledge signals at coarse level.  A higher value of Jaccard Index indicates that very few sentences have inched into the personal summary, and the algorithm probably did not adequately acknowledge the user preferences. Lower value of JI indicates the effectiveness of the algorithm, that is, new sentences are there in $S_p$ due to either exclusion of knowledge semantically related to $\mathcal K^- $ or inclusion of more knowledge semantically similar to  $\mathcal K^+$. Though lowering of JI connotes personalization of summary, we expect a steeper fall in the metric value for $\mathcal K^- $ compared to $\mathcal K^+ $ because of retention of sentences with positive signals in $S_g$. This is clearly indicated in our experiments.} 

\item{Term-level Evaluation:}
\label{subsec:term-level-evaluation}
 \new{Considering $S_p$ and $S_g$ as collections of terms, the difference in their distributions indicates the extent of new content included in $S_p$ compared to $S_g$. The distance between distributions of terms in the two summaries is proportional to the amount of new knowledge injected in the personal summary relative to the generic summary. Recall that for every term in $\mathcal K$, we consider all co-occurring terms as semantically related and assign weight appropriately in matrix $W$ (Sec. \ref{sec-weight-matrix}). Therefore, we expect that the new content in the \textit{personal} matches the user knowledge preference.}

\new{We employ Jensen-Shannon distance to quantify the difference between distributions of terms in \textit{personal} and \textit{generic} summaries for quality assessment of $S_p$. The high value of the metric indicates admittance of higher number of new terms, which proxy for new knowledge in the personal summary. With the increase in number of terms in knowledge signals, distance between the two distributions is expected to increase.}   

\item{Semantic-level Evaluation:}
\label{item:semantic-level-evaluation}
\new{Sentence- and term-level measures have lexical orientation and are inadequate to assess how a personal summary is \textit{semantically} different from the generic summary. Semantic-level evaluation is the confirmatory test that reveals how well the summary acknowledges user preferences. In other words, quantifying the difference between $S_p$ and $S_g$ at the latent semantic level reveals ability of the algorithm to personalize at the finest level of granularity. We perform this evaluation by comparing the semantic similarity between the knowledge signals and, $S_p$ and $S_g$ as described below.} 

\new{Let $\sigma(S, \mathcal K)$ denote the semantic similarity between knowledge signals and the summary\footnote{We present the method of computing semantic similarity in Appendix \ref{appendix-semantic-similarity}.}.  Under the assumption that $S_g$ comprises \textit{important sentences} about the \textit{salient} \textit{topics} contained in $D$, we conjecture that personal summary, $S_p$ is semantically less similar to negative knowledge signals than the generic summary. Mathematically,   $\sigma(S_p, \mathcal K^-) \le \sigma(S_g, \mathcal K^-)$. The underlying reason is that $\mathcal K^-$ has been stripped off from  $S_p$  by the algorithm, while $S_g$ may contain it in bits-and-pieces. On the other hand, for positive (desired) knowledge signals, the personal summary is richer in  $\mathcal K^+$ than the generic summary,  that is, $\sigma(S_p, \mathcal K^+) \ge \sigma(S_g, \mathcal K^+)$. }

\new{We quantify the effectiveness of personal summaries at the semantic level by computing the ratio of semantic similarity of knowledge signal with personal and generic summaries, and denote it as follows. }
\begin{equation}
	\mathcal R =\frac{\sigma(S_p,\mathcal K)}{\sigma(S_g,\mathcal K)}  
	\label{eqn-ratio-summ-knowledge}
\end{equation}
\new{  For clarity, we denote semantic similarity of the summaries with negative knowledge signals by $\mathcal R^-$, and with positive knowledge signals as $\mathcal R^+$. Ideally, the personal summary that acknowledges negative knowledge signals results in $\mathcal R^- \leq 1$, and the one for positive signals results in $ \mathcal R^+ \geq 1$.}

\new{The premise stated for $\mathcal R^-$ and $\mathcal R^+$ may not consistently hold as their values might deviate from the ideal range. This is because an extractive summary is influenced by several factors, including the writing style of document, its length, and the quality of keywords used as knowledge signals. Furthermore, for the combination of the set of keywords used as user knowledge signals and the length of the personal summary, ratio $\mathcal R^-$ should ideally be less than the corresponding value of $\mathcal R^+$. This is because a personal summary that respects $\mathcal K^-$ suppresses information in personal summary and is thus semantically less close to $S_g$ than the personal summary generated by considering the same signals as $\mathcal K^+$. }
\end{enumerate}
\new{\textit{The use of a single system-generated summary as the gold standard and the combination of three tests in lexical and latent semantic space of the document makes the framework practicable and robust. Thus, the framework addresses the protocol and metric issues in an elegant, effective, and efficient manner.}}
\section{Experimental Design}
\label{sec:experimental-design}
In this section, we describe the experimental setup for performance evaluation of the proposed \textit{P-Summ} algorithm. The algorithm is implemented in Python 3.7 using \textit{NLTK}, \textit{scikit-learn}, \textit{textmining},  and \textit{wNMF} libraries. All the experiments are performed on Windows 10 machine with Intel(R) Core(TM) i7-10700 CPU and 64 GB RAM. 

\new{We consciously decide to omit comparative evaluation of the proposed algorithm with previous works in the area of personalized summarization. Existing works, which use\textit{ aspect}, \textit{query} or \textit{entity} as mode of knowledge signal cannot be directly compared with \textit{P-Summ} due to incompatibility of the benchmark datasets used for evaluation  (Please see item \ref{challenge-ii} of Sec. \ref{subsec:research-gap}). We examined three datasets for aspect-based summarization  \citep{hayashi2021wikiasp, ahuja2022aspectnews, yang2022oasum}, and realized that it is not straightforward to map the \textit{abstract} aspects (e.g. sentiment, geography, culture, history, etc.) to the terms in an article. Therefore, we maintain the focus on evaluating the effectiveness of \textit{P-Summ} algorithm to eliminate undesired knowledge corresponding to the negative knowledge signals and introduce desired knowledge for positive signals. }
\begin{table}[ht]
	\begin{minipage}[b]{\linewidth}
		\centering
        \caption{Overview of experimental datasets. $|D|$: Number of documents, $L_{avg}$: Average document length (in words), $N_{avg}$: Average number of sentences, $S_{avg}$: Average sentence length (in words) }
		\begin{tabular}{l|c|c|c|c}
			\hline
			& \multicolumn{3}{c|}{\textbf{CL-SciSumm Test Set}} &  \\ \cline{2-4}
			& \multicolumn{1}{l|}{\textbf{2016}} & \multicolumn{1}{l|}{\textbf{2017}} & \textbf{2018} &\textbf{AIPubSumm Test Set} \\ \hline
			Domain& \multicolumn{3}{c|}{Computational Linguistic} & Artificial Intelligence \\ \hline
			$|D|$ & \multicolumn{1}{l|}{10} & \multicolumn{1}{l|}{10} & 20 & 62 \\ \hline
			$L_{avg}$ & \multicolumn{1}{l|}{3505} & \multicolumn{1}{l|}{2954} & 3467 & 6736 \\ \hline
			$N_{avg}$ & \multicolumn{1}{l|}{175} & \multicolumn{1}{l|}{158} & 165 & 316 \\ \hline
			$S_{avg}$ & \multicolumn{1}{l|}{20} & \multicolumn{1}{l|}{19} & 21 & 21 \\ \hline
		\end{tabular}
		\label{table-datasets}
	\end{minipage}
\end{table}

\noindent \textbf{Datasets:}
We use four public datasets comprising of scientific scholarly articles to assess the quality of \textit{personal} summaries generated by \textit{P-Summ} algorithm. The first three datasets are  collections of ACL articles from  CL-SciSumm Shared Task datasets 2016 -  2018 \citep{scisumm2016overview, scisumm2017overview, scisumm2018overview}. The fourth dataset,   AIPubSumm,  consists of research articles from Artificial Intelligence domain \citep{AIPubSum-dataset}.  Since \textit{P-Summ} is an unsupervised algorithm, we do not use training and validation sets and evaluate the performance of the proposed algorithm using test sets. Table \ref{table-datasets} summarizes the basic statistics for the four experimental datasets.\\

\noindent \textbf{Simulating User Knowledge Signal:}
\new{We consider author-defined keywords as prominent topics or concepts described in a scientific article and use them to simulate user knowledge signals for generating personal summaries. Each article in AIPubSumm dataset is accompanied by a set of at most five author-defined keywords\footnote{We consider 62/66 documents containing author-defined keywords for experimentation. }, which proxy as $\mathcal K^-$ or $\mathcal K^+$. Since author-defined keywords for CL-SciSumm datasets are not available, we extract five top-scoring bi-grams from the documents using BERT-based keyword extraction algorithm, KeyBERT \citep{grootendorst2020keybert}, and use them to simulate user knowledge preferences.} \\

\noindent \textbf{Evaluation Strategy:}
\new{Our experiments span over three dimensions along which the proposed \textit{P-Summ} algorithm is evaluated.}
\begin{enumerate}[i. ]
    \item \new{ Type of knowledge signal - We use the same set of keywords as negative and positive knowledge signals and generate corresponding personal summaries. Comparison of evaluation metrics for the two contrasting summaries highlights the effectiveness of scoring functions in Equations \ref{eqn-penalty} and \ref{eqn-reward}, and the proposed \textit{P-Summ} algorithm. }

    \item \new{Strength of knowledge signal - The strength of knowledge signal is controlled by the number of keywords comprising it. We generate summaries for each knowledge type ($\mathcal K^-$ or $\mathcal K^+$) and signal strength varying from 1 to 5 keywords. Quantitative evaluation of these summaries reveals the effect of summary length on the level of personalization achieved by the algorithm. }

    \item \new{ Summary Length - We investigate the effect of increasing summary lengths on the level of personalization achieved by the proposed algorithm.}
\end{enumerate}

\noindent \textbf{Presentation of Results:}
\new{We present the results for each dataset in three tables corresponding to the three metrics viz. Jaccard Index, Jensen Shannon Distance and ratio $\mathcal R$, described in  Sec. \ref{sec:evaluation-framework}. Each table is horizontally partitioned into two parts and shows the results for negative and positive signals for the same set of keywords. The columns in the two partitions correspond to different summary lengths. The four columns of the left partition present the results for negative knowledge signals, and those on right show the results for positive knowledge signals. }

\new{The table consists of five rows corresponding to the number of keywords, which signify the strength of knowledge signals $\mathcal K$ ($\mathcal K^-$ or $\mathcal K^+$). Each value in the cell is macro-averaged over the algorithmic personal summaries of specific summary length and signal strength for all the documents in the dataset. The metric score for a document in the dataset is computed by averaging the scores of the personal summaries obtained by considering all possible combinations of keywords in $\mathcal K$.  The following example clarifies the computation of score for a document for a specific summary length and signal strength. }
\begin{Ex} 
Consider a document $D$ with five author-specified keywords. Suppose we choose to compute quality score for the summary of length $l$ and signal strength of three keywords. We generate $5 \choose 3$ personal summaries of length $l$ corresponding to all possible combinations of keywords and compute evaluation metrics for each summary. Subsequently, we average $5 \choose 3$  scores to assess summary quality for $D$ with knowledge signal comprising three keywords.  
\end{Ex}  
\noindent
\new{Averaging the metric score over all possible combinations of keywords mitigates the effect of variations due to the quality of keywords. Our evaluation strategy implicitly takes into account the quality of keywords and comprehensively evaluates the effect of type, strength of knowledge signals, and summary length on the quality of the personal summary.}

\section{Performance Evaluation of P-Summ}
\label{sec:performance-evaluation-psumm}
This section reports the results of performance evaluation of the proposed \textit{P-Summ} algorithm using four datasets described in Sec. \ref{sec:experimental-design}.  It is challenging to evaluate the efficacy of the proposed \textit{P-Summ} algorithm while considering negative and positive knowledge preferences simultaneously. \new{The primary reason is that there is a possibility that the gold standard reference summary may not reflect the specified combination of both positive and negative knowledge signals. Human evaluation of personal summary is also subjective and unreliable due to the inherent ability of the human brain to fill gaps and omit terms while reading.}. Therefore, we assess the performance of \textit{P-Summ} algorithm while considering negative and positive knowledge preferences separately. We present dataset-wise results in Sections \ref{subsec:perf-evaluation-AIPubSumm} (Table \ref{table-AIPubSumm}) and \ref{subsec:perf-evaluation-CLSciSumm} (Tables \ref{table-CLSciSumm2016}-\ref{table-CLscisumm2018}). 

\subsection{Performance Evaluation on AIPubSumm dataset}
\label{subsec:perf-evaluation-AIPubSumm}
Tables \ref{table-AIPubSumm-Jaccard} - \ref{table-AIPubSumm-SemanticSimilarity} present the results for personal summaries for the articles in AIPubSumm test set, where author-defined keywords given with the articles serve as knowledge signals.
\begin{table}[h]
	\centering
	\scriptsize
    \caption{Performance of \textit{P-Summ} on AIPubSumm dataset. kw: signal strength (number of keywords in $\mathcal K$), $\mathcal D$: document length, JI: Jaccard Index, JSD: Jensen-Shannon distance, $\mathcal R$: ratio of semantic similarities (Eq. \ref{eqn-ratio-summ-knowledge})}
	\begin{subtable}[t]{\linewidth}
		\centering
        \caption{Jaccard Index for AIPubSumm dataset}
		\resizebox{\textwidth}{!}{\begin{tabular}{c|c|c|c|c||c|c|c|c}
			\hline
			&\multicolumn{4}{c||}{Negative Signals}&\multicolumn{4}{c}{Positive Signals}\\
			\hline
			kw&{$10\%$ * $\mathcal{D}$}&{$15\%$ * $\mathcal{D}$}&{$20\%$ * $\mathcal{D}$}&{$25\%$ * $\mathcal{D}$}	&{$10\%$ * $\mathcal{D}$}&{$15\%$ * $\mathcal{D}$}&{$20\%$ * $\mathcal{D}$}&{$25\%$ * $\mathcal{D}$}\\
			\hline
			1 &0.1782	&0.1850	&0.1907	&0.2020	&0.7025	&0.7305	&0.7190	&0.7384\\
			\hline
			2 &0.1021	&0.1031	&0.1114	&0.1218	&0.7045	&0.7249	&0.7204	&0.7436\\
			\hline
			3 &0.0853	&0.0793	&0.0889	&0.0983	&0.7105	&0.7243	&0.7168	&0.7445\\
			\hline
			4 &0.0892	&0.0837	&0.0862	&0.0931	&0.7121	&0.7223	&0.7101	&0.7461\\
			\hline
			5 &0.0966	&0.0896	&0.0917	&0.0919	&0.6978	&0.7182	&0.7356	&0.7604\\
			\hline		
		\end{tabular}}
		
		\label{table-AIPubSumm-Jaccard}
		\quad
	\end{subtable}
	\\
        \vspace{2mm}
	\begin{subtable}[t]{\linewidth}
		\centering
        \caption{Jensen-Shannon Distance for AIPubSumm dataset }
		\resizebox{\textwidth}{!}{\begin{tabular}{c|c|c|c|c||c|c|c|c}
			\hline
			&\multicolumn{4}{c||}{Negative Signals}&\multicolumn{4}{c}{Positive Signals}\\
			\hline
			&{$10\%$ * $\mathcal{D}$}&{$15\%$ * $\mathcal{D}$}&{$20\%$ * $\mathcal{D}$}&{$25\%$ * $\mathcal{D}$}	&{$10\%$ * $\mathcal{D}$}&{$15\%$ * $\mathcal{D}$}&{$20\%$ * $\mathcal{D}$}&{$25\%$ * $\mathcal{D}$}\\
			\hline
			1 &0.7007	&0.6695	&0.6410	&0.6167	&0.3086	&0.2734	&0.2700	&0.2522\\
			\hline
			2 &0.7856	&0.7504	&0.7168	&0.6886	&0.3014	&0.2775	&0.2683	&0.2482\\
			\hline
			3 &0.8151	&0.7766	&0.7399	&0.7120	&0.3039	&0.2856	&0.2717	&0.2486\\
			\hline
			4 &0.8219	&0.7795	&0.7449	&0.7149	&0.3144	&0.2912	&0.2837	&0.2473\\
			\hline
			5 &0.8203	&0.7799	&0.7486	&0.7240	&0.3258	&0.3058	&0.2690	&0.2374\\
			\hline
		\end{tabular}}
		
		\label{table-AIPubSumm-JSD}
		\quad
	\end{subtable}
	\\
        \vspace{2mm}
	\begin{subtable}[t]{\linewidth}
		\centering
        \caption{ Ratio $\mathcal{R}$ for AIPubSumm dataset}
		\resizebox{\textwidth}{!}{\begin{tabular}{c|c|c|c|c||c|c|c|c}
			\hline
			&\multicolumn{4}{c||}{Negative Signals ($\mathcal R^-$)}&\multicolumn{4}{c}{Positive Signals ($\mathcal R^+$)}\\
			\hline
			&{$10\%$ * $\mathcal{D}$}&{$15\%$ * $\mathcal{D}$}&{$20\%$ * $\mathcal{D}$}&{$25\%$ * $\mathcal{D}$}	&{$10\%$ * $\mathcal{D}$}&{$15\%$ * $\mathcal{D}$}&{$20\%$ * $\mathcal{D}$}&{$25\%$ * $\mathcal{D}$}\\
			\hline
			1 &0.7375	&0.7496	&0.7597	&0.7624	&1.0127	&1.0140	&1.0166	&0.9966\\
			\hline
			2 &0.7617	&0.7632	&0.7656	&0.7658	&1.0113	&1.0103	&1.0134	&1.0017\\
			\hline
			3 &0.8328	&0.8234	&0.8139	&0.8206	&1.0106	&1.0024	&1.0025	&0.9989\\
			\hline
			4 &0.8254	&0.7950	&0.7836	&0.7913	&1.0016	&0.9896 &0.9918	&0.9920\\
			\hline
			5 &0.8504	&0.8200	&0.8243	&0.8095	&1.0127	&1.0021	&1.0032	&0.9936\\
			\hline
		\end{tabular}}
		
		\label{table-AIPubSumm-SemanticSimilarity}
		\quad
	\end{subtable}
	\label{table-AIPubSumm}
\end{table}

\subsubsection{Results for Negative Knowledge Signals}
\label{subsubsec:perf-evaluation-AIPubSumm-negative}
Table \ref{table-AIPubSumm-Jaccard} shows Jaccard Index (JI)  for different summary lengths (in columns) and signal strengths (in rows). A decrease in the number of overlapping sentences with an increase in the signal strength is attributed to the greater elimination of the user-specified knowledge and induction of \textit{new} knowledge in the personal summary. As the number of keywords in $\mathcal K^-$ increases, the summary becomes increasingly personalized and the commonality between the personal summary ($S_p$) and generic summary ($S_g$)  decreases, thereby lowering the Jaccard Index. 

Consistently increasing Jensen-Shannon distance with increasing signal strength ratifies the trend exhibited by JI (see first four columns of Table \ref{table-AIPubSumm-JSD}). Induction of new sentences due to increased signal strength (i.e. number of keywords) leads to divergence in the distribution of terms in $S_p$ and $S_g$, which manifests as increasing Jensen-Shannon distance.

\begin{figure}[ht]
\centering
\includegraphics[width=0.95\textwidth]{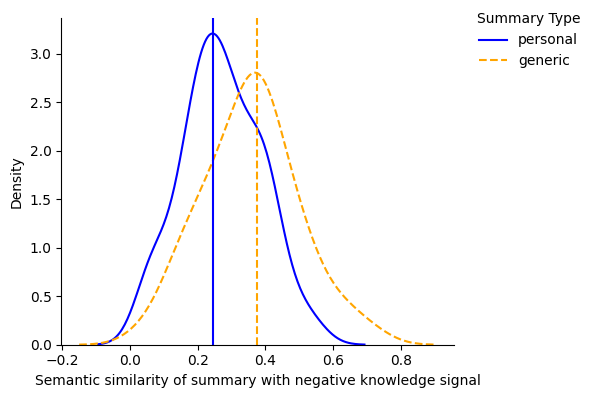}
\caption{Density plot for semantic similarities between  $\mathcal {K^-}$ and,  personal and generic summaries for AIPubSumm dataset for 25\% summary length and one keyword.}
\label{fig-AI-density-plot-neg}
\end{figure}

Figure \ref{fig-AI-density-plot-neg} displays density plot of the semantic similarity values of \textit{personal} and \textit{generic} summaries for all documents for negative knowledge and one keyword. The length of personal and generic summary is 25\% of the document length. The peak of the density curve for personal summaries is towards the left of the peak for generic summaries, indicating that the personal summaries for most documents have lower semantic similarities with knowledge signal compared to corresponding generic summaries. The plot also shows that there is more variability in the similarities of personal and generic summaries, as the tail of density curves for generic summaries extends towards the right.

Finally, the ratio of semantic similarities between $\mathcal K^-$ and $S_p$, and $\mathcal K^-$ and $S_g$ (Eq. \ref{eqn-ratio-summ-knowledge}) confirms that the personal summaries are semantically distant from  $\mathcal K^-$ than the generic summaries (first four columns of Table \ref{table-AIPubSumm-SemanticSimilarity}). Personal summaries respect the user knowledge needs and bear lesser content related to $\mathcal K^-$ with increasing signal strength.

When the performance metrics are observed across rows for $\mathcal K^-$ in Tables \ref{table-AIPubSumm-Jaccard} and \ref{table-AIPubSumm-JSD}, it is noted that the degree of personalization diminishes with increasing summary length for a specific signal strength. Increasing  Jaccard Index and decreasing  Jensen-Shannon distance both indicate that it is difficult to \textit{hold-back} negative knowledge for increasing summary lengths. Long personal summaries exhibit a larger overlap of sentences with generic summaries. Some important sentences that were \textit{down-scored} earlier creep into personal summary when summary length increases, causing a larger overlap of generic summaries. 

\subsubsection{Results for Positive Knowledge Signals}
\label{subsubsec:perf-evaluation-AIPubSumm-positive}
Columns on the right, in Table \ref{table-AIPubSumm-Jaccard} - \ref{table-AIPubSumm-SemanticSimilarity}, present the results of evaluation of \textit{P-Summ} on AIPubSumm dataset for positive knowledge signals. For each combination of signal strength and summary length, persistently higher values of the Jaccard Index for $\mathcal K^+$ compared to the corresponding entries for $\mathcal K^-$ suggest that the personal summary corresponding to positive knowledge decisively ensconces the information desired by the user (Table \ref{table-AIPubSumm-Jaccard}, right four columns).  \textit{We note that the JI values do not change significantly with the increase in signal strength for all summary lengths. This is because a generic summary already includes sentences containing the specified positive signals, and rewarding these keywords in weight matrix $W$ does not bring significant change in the importance of sentences containing these keywords.
}

A similar trend is evident for the values of Jensen-Shannon distance for negative and positive knowledge signals for all signal strengths (Table \ref{table-AIPubSumm-JSD}). This asserts that when the same set of keywords is used for simulating both types of knowledge signals, \textit{P-Summ} algorithm assertively includes the terms desired by the user and omits the unwanted keywords and other related terms. Additionally, the distribution of terms in the personal summary for $\mathcal K^+$ is closely aligned with that of the generic summary ($S_g$).

\begin{figure}[ht]
\centering
\includegraphics[width=0.95\textwidth]{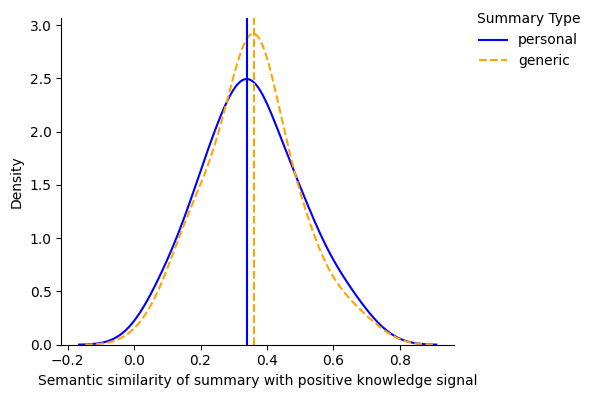}
\caption{Density plot for semantic similarities between  $\mathcal {K^+}$ and,  personal and generic summaries for AIPubSumm dataset}
\label{fig-AI-density-plot-pos}
\end{figure}

Figure \ref{fig-AI-density-plot-pos} displays density plot for the similarities of personal and generic summaries with positive knowledge signal with one keyword and summary length of 25\%. Approximately overlapping peaks of the density curves for personal and generic summaries indicate commonality in these summaries. This is the expected behaviour for positive knowledge signals because the generic summary is already an effective representation of important content and includes sentences containing the keywords specified as $\mathcal K^+$.

Table \ref{table-AIPubSumm-SemanticSimilarity} for positive knowledge signals shows that ratios $\mathcal R^+$ are conclusively higher than  $\mathcal R^-$.  The premise stated in item \ref{item:semantic-level-evaluation} of Sec.~\ref{subsec:eval-metric}  holds prominently for combinations of summary length and signal strength that admit more positive knowledge. As expected, for lower signal strength and smaller summary length, ratio $\mathcal R^+$ deviates slightly from the ideal behavior.  This is because short summaries for $\mathcal K^+$ are unable to garner adequate positive knowledge. It is noteworthy that the quality of keywords used as knowledge signal also influences the performance of the algorithm.

\subsection{Results for CL-SciSumm datasets}
\label{subsec:perf-evaluation-CLSciSumm}
We present the results of the performance evaluation of \textit{P-Summ} algorithm over CL-SciSumm 2016 - 2018  datasets in this section. The evaluation protocol for CL-SciSumm datasets was same as that of AIPubSumm dataset, except that the keywords used as knowledge signals are extracted 
 using KeyBERT algorithm \citep{grootendorst2020keybert}.

\subsubsection{Results for CL-SciSumm 2016 dataset}
\label{subsubsec:perf-evaluation-CLSciSumm2016}

\begin{table}[ht]
        \scriptsize
	\centering
         \caption{Performance of \textit{P-Summ} on CL-SciSumm 2016 dataset. kw: signal strength (number of keywords in $\mathcal K$), $\mathcal D$: document length, JI: Jaccard Index, JSD: Jensen-Shannon distance, $\mathcal R$: ratio of semantic similarities (Eq. \ref{eqn-ratio-summ-knowledge})}
	\begin{subtable}[t]{\linewidth}
		\centering
        \caption{Jaccard Index for CL-SciSumm 2016 dataset }
		\resizebox{\textwidth}{!}{\begin{tabular}{c|c|c|c|c||c|c|c|c}
			\hline
			&\multicolumn{4}{c||}{Negative  Signals}&\multicolumn{4}{c}{Positive Signals}\\
			\hline
			kw&{$10\%$ * $\mathcal{D}$}&{$15\%$ * $\mathcal{D}$}&{$20\%$ * $\mathcal{D}$}&{$25\%$ * $\mathcal{D}$}	&{$10\%$ * $\mathcal{D}$}&{$15\%$ * $\mathcal{D}$}&{$20\%$ * $\mathcal{D}$}&{$25\%$ * $\mathcal{D}$}\\
			\hline
	           1 &0.3804	&0.3786	&0.3850	&0.3990&0.7190	&0.6772	&0.6759	&0.7309\\
			\hline
			2 &0.2426	&0.2349	&0.2463	&0.2459&0.6989	&0.6942	&0.6767	&0.7230\\
			\hline
			3 &0.1736	&0.1703	&0.1881	&0.1839&0.6938	&0.7020	&0.6762	&0.7271\\
			\hline
			4 &0.1379	&0.1336	&0.1570	&0.1545&0.6915	&0.7065	&0.6786	&0.7322\\
			\hline
			5 &0.1064	&0.1122	&0.1422	&0.1417&0.7100	&0.7048	&0.6727	&0.7285\\
			\hline		
		\end{tabular}}
		\label{table-CLSciSumm2016-Jaccard}
		\quad
	\end{subtable}
	\\
        \vspace{2mm}
	\begin{subtable}[t]{\linewidth}
		\centering
        \caption{Jensen-Shannon Distance for CL-SciSumm 2016 dataset}
		\resizebox{\textwidth}{!}{\begin{tabular}{c|c|c|c|c||c|c|c|c}
			\hline
			&\multicolumn{4}{c||}{Negative Signals}&\multicolumn{4}{c}{Positive Signals}\\
			\hline
			kw&{$10\%$ * $\mathcal{D}$}&{$15\%$ * $\mathcal{D}$}&{$20\%$ * $\mathcal{D}$}&{$25\%$ * $\mathcal{D}$}	&{$10\%$ * $\mathcal{D}$}&{$15\%$ * $\mathcal{D}$}&{$20\%$ * $\mathcal{D}$}&{$25\%$ * $\mathcal{D}$}\\
			\hline
			 1 &0.5151	&0.5146	&0.5039	&0.4840	&0.2592	&0.2756	&0.2808	&0.2411\\
			\hline
			  2 &0.6414	&0.6402	&0.6216	&0.6049	&0.2756	&0.2686	&0.2779	&0.2470\\
			\hline
			3 &0.7090	&0.7021	&0.6774	&0.6557	&0.2839	&0.2622	&0.2760	&0.2426\\
			\hline
			4 &0.7465	&0.7371	&0.7071	&0.6802	&0.2863	&0.2580	&0.2718	&0.2398\\
			\hline
			 5 &0.7735	&0.7589	&0.7211	&0.6924	&0.2691	&0.2573	&0.2756	&0.2457\\
			\hline
		\end{tabular}}
		\label{table-CLSciSumm2016-JSD}
		\quad
	\end{subtable}
        \\
        \vspace{2mm}
	\begin{subtable}[t]{\linewidth}
		\centering
        \caption{Ratio $\mathcal{R}$ for CL-SciSumm 2016 dataset}
		\resizebox{\textwidth}{!}{\begin{tabular}{c|c|c|c|c||c|c|c|c}
			\hline
			&\multicolumn{4}{c||}{Negative Signals ($\mathcal R^-$)}&\multicolumn{4}{c}{Positive Signals ($\mathcal R^+$)}\\
			\hline
			kw&{$10\%$ * $\mathcal{D}$}&{$15\%$ * $\mathcal{D}$}&{$20\%$ * $\mathcal{D}$}&{$25\%$ * $\mathcal{D}$}	&{$10\%$ * $\mathcal{D}$}&{$15\%$ * $\mathcal{D}$}&{$20\%$ * $\mathcal{D}$}&{$25\%$ * $\mathcal{D}$}\\
			\hline
			1 &0.8253	&0.8233	&0.9209	&0.8765	&0.9052	&0.9624	&1.0288	&0.9662\\
			\hline
			2 &0.8409	&0.7541	&0.8350	&0.7776	&0.9261	&0.9615	&1.0347	&0.9673\\
			\hline
			3 &0.8378	&0.8070	&0.8444	&0.8072	&0.9257	&1.0013	&1.0577	&0.9868\\
			\hline
			4 &0.8895	&0.8671	&0.9050	&0.8805	&0.9382	&1.0118	&1.0690	&1.0028\\
			\hline
			5 &0.8866	&0.8731	&0.9181	&0.8929	&0.9624	&1.0272	&1.0843	&1.0148\\
			\hline
		\end{tabular}}
		\label{table-CLSciSumm2016-SemanticSimilarity}
		\quad
	\end{subtable}
	\label{table-CLSciSumm2016}
\end{table}

Tables \ref{table-CLSciSumm2016-Jaccard} - \ref{table-CLSciSumm2016-SemanticSimilarity} display the results of quality assessment of personal summaries for the articles in CL-SciSumm 2016 dataset. Here also, we notice that the trends agree with the earlier observations for all three metrics. A considerable gap in the metric values for positive and negative knowledge signals is evident in Tables \ref{table-CLSciSumm2016-Jaccard} - \ref{table-CLSciSumm2016-SemanticSimilarity}.  We consolidate below the trends observed for CL-SciSumm 2016 dataset.
\begin{enumerate}[i.]
    \item Jaccard Index (Table \ref{table-CLSciSumm2016-Jaccard}) \\
    a) In each column for $\mathcal K^-$, scores decrease with increase in signal strength, indicating progressive elimination of \textit{unwanted} knowledge from the personal summary ($S_p$). The trend holds for all tested summary lengths.\\
    b) Along the columns for $\mathcal K^+$, scores don't change much as the number of keywords increases. The observation holds for all summary lengths. This is because the personal summary for \textit{positive} knowledge signals is very similar to the generic summary ($S_g$), which inherently comprises the important sentences encompassing the information conveyed by the keywords.\\
   
    \item Jensen-Shannon Distance (Table \ref{table-CLSciSumm2016-JSD})\\
    a) Jensen-Shannon Distance (JSD) increases for $\mathcal K^-$ with increase in the signal strength (column-wise), indicating elimination of \textit{known} knowledge from the personal summary and more deviation from the generic summary at the term level. The trend remains nearly uniform for all tested summary lengths (row-wise).\\
    b) JSD does not deviate much for $\mathcal K^+$ as the number of keywords increase (column-wise) for all summary lengths, which indicates the inclusion of fewer new terms in the personal summary as compared to the generic summary possessing the information reflected in the keywords used for simulating the user preferences.\\
    
    \item Ratios (Table \ref{table-CLSciSumm2016-SemanticSimilarity})\\
    a) Values for ratio $\mathcal R^+$ (right four columns) are higher than the corresponding entries for $\mathcal R^-$ (left four columns), because the proposed \textit{P-Summ} algorithm effectively eliminates \textit{known} knowledge to create a personal summary for the given $\mathcal K^-$, and adds the desired knowledge in the personal summary for $\mathcal K^+$.
    
\end{enumerate}

\label{subsubsec:perf-comparison-CLSciSumm-negative-positive}
\begin{figure}[h]
\centering
\begin{subfigure}{0.48\textwidth}
\centering
\includegraphics[width=\textwidth]{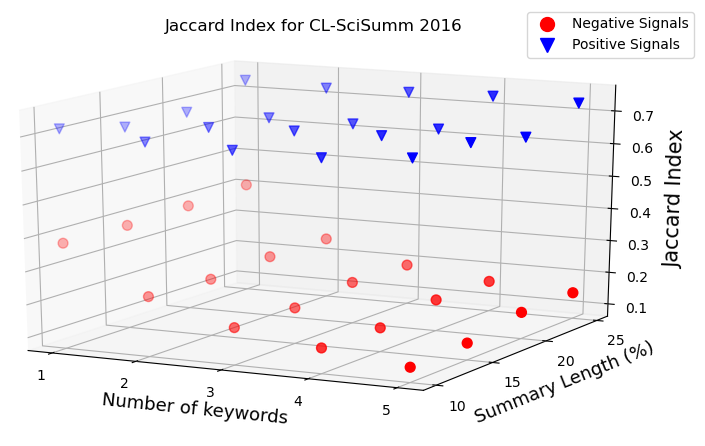}
\caption{Jaccard Index}
\label{fig-scatter-scisumm16-jaccard}
\end{subfigure}
\begin{subfigure}{0.48\textwidth}
\centering
\includegraphics[width=\textwidth]{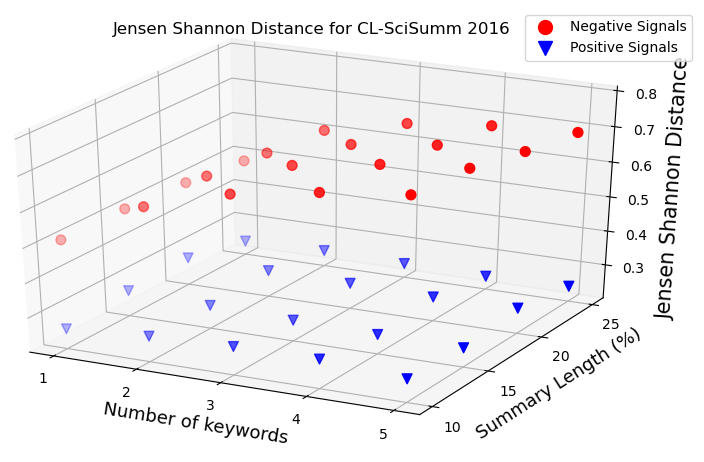}
\caption{Jensen Shannon Distance}
\label{fig-scatter-scisumm16-jsd}
\end{subfigure}
\caption{Comparison of performance for negative and positive knowledge signals in CL-SciSumm 2016 dataset based on Jaccard Index and Jensen Shannon Distance metrics}
\label{fig-scatter-scisumm16}
\end{figure}

Plots in Figures \ref{fig-scatter-scisumm16-jaccard} and \ref{fig-scatter-scisumm16-jsd} present comparative performance of \textit{P-Summ} for positive and negative knowledge signals on CL-SciSumm 2016  documents using Jaccard Index and Jensen Shannon Distance, respectively. The clear separation of the data points for positive and negative knowledge signals reveals the efficacy of the proposed algorithm in generating user-specific summaries. The algorithm adequately eliminates the knowledge \textit{not desired} by the user ($\mathcal K^-$) and incorporates the \textit{desired} knowledge in the personal summary, and consequently, Jaccard values for $\mathcal K^-$ are noticeably lower than those for $\mathcal K^+$. Jensen-Shannon Distance also exhibits a contrasting trend demonstrating the efficacy of the proposed algorithm at term level.  The adherence of all three evaluation metrics to the hypothesized trend affirms the effectiveness of \textit{P-Summ} algorithm. 

\begin{table}[ht]
	\centering
        \caption{Performance of \textit{P-Summ} on CL-SciSumm 2017 dataset. kw: signal strength (number of keywords in $\mathcal K$), $\mathcal D$: document length, JI: Jaccard Index, JSD: Jensen-Shannon distance, $\mathcal R$: ratio of semantic similarities (Eq. \ref{eqn-ratio-summ-knowledge})}
	\scriptsize
	\begin{subtable}[t]{\linewidth}
		\centering
        \caption{Jaccard Index for CL-SciSumm 2017 dataset}
		\resizebox{\textwidth}{!}{\begin{tabular}{c|c|c|c|c||c|c|c|c}
			\hline
			&\multicolumn{4}{c||}{Negative Signals}&\multicolumn{4}{c}{Positive Signals}\\
			\hline
			kw&{$10\%$ * $\mathcal{D}$}&{$15\%$ * $\mathcal{D}$}&{$20\%$ * $\mathcal{D}$}&{$25\%$ * $\mathcal{D}$}	&{$10\%$ * $\mathcal{D}$}&{$15\%$ * $\mathcal{D}$}&{$20\%$ * $\mathcal{D}$}&{$25\%$ * $\mathcal{D}$}\\
			\hline
			1 &0.3114	&0.2957	&0.3471	&0.3663	&0.6527	&0.6603	&0.7227	&0.7261\\
			\hline
			2 &0.1784	&0.1724	&0.2258	&0.2517	&0.6325	&0.6395	&0.7176	&0.7223\\
			\hline
			3 &0.1254	&0.1273	&0.1700	&0.1927	&0.6313	&0.6356	&0.7195	&0.727\\
			\hline
			4 &0.0983	&0.1032	&0.1407	&0.1585	&0.6311	&0.6317	&0.7158	&0.7248\\
			\hline
			5 &0.0889	&0.0929	&0.1168	&0.1315	&0.6207	&0.6322	&0.7122	&0.7209\\
			\hline		
		\end{tabular}}
		\label{table-CLSciSumm2017-Jaccard}
		\quad
	\end{subtable}
	\\
        \vspace{2mm}
	\begin{subtable}[t]{\linewidth}
		\centering
        \caption{Jensen-Shannon Distance CL-SciSumm 2017 dataset}
		\resizebox{\textwidth}{!}{\begin{tabular}{c|c|c|c|c||c|c|c|c}
			\hline
			&\multicolumn{4}{c||}{Negative Signals}&\multicolumn{4}{c}{Positive Signals}\\
			\hline
			kw&{$10\%$ * $\mathcal{D}$}&{$15\%$ * $\mathcal{D}$}&{$20\%$ * $\mathcal{D}$}&{$25\%$ * $\mathcal{D}$}	&{$10\%$ * $\mathcal{D}$}&{$15\%$ * $\mathcal{D}$}&{$20\%$ * $\mathcal{D}$}&{$25\%$ * $\mathcal{D}$}\\
			\hline
			1 &0.6292	&0.5987	&0.5535	&0.5256	&0.3356	&0.3341	&0.2623	&0.2601\\
			\hline
			2 &0.7393	&0.6933	&0.6378	&0.6077	&0.3625	&0.3582	&0.2730	&0.2605\\
			\hline
			3 &0.7849	&0.7353	&0.6831	&0.6551	&0.3648	&0.3661	&0.2737	&0.2556\\
			\hline
			4 &0.8085	&0.7624	&0.7089	&0.6831	&0.3648	&0.3715	&0.2793	&0.2595\\
			\hline
			5 &0.8211	&0.7803	&0.7316	&0.7029	&0.3799	&0.3780	&0.2864	&0.2642\\
			\hline	
		\end{tabular}}
		\label{table-CLSciSumm2017-JSD}
		\quad
	\end{subtable}
	\\
        \vspace{2mm}
	\begin{subtable}[t]{\linewidth}
		\centering
        \caption{Ratio $\mathcal{R}$ for CL-SciSumm 2017 dataset}
		\resizebox{\textwidth}{!}{\begin{tabular}{c|c|c|c|c||c|c|c|c}
			\hline
			&\multicolumn{4}{c||}{Negative Signals ($\mathcal R^-$)}&\multicolumn{4}{c}{Positive Signals ($\mathcal R^+$)}\\
			\hline
			kw&{$10\%$ * $\mathcal{D}$}&{$15\%$ * $\mathcal{D}$}&{$20\%$ * $\mathcal{D}$}&{$25\%$ * $\mathcal{D}$}	&{$10\%$ * $\mathcal{D}$}&{$15\%$ * $\mathcal{D}$}&{$20\%$ * $\mathcal{D}$}&{$25\%$ * $\mathcal{D}$}\\
			\hline
			1 &0.8238	&0.8727	&0.9630	&0.8835	&0.9626	&1.0497	&1.1129	&1.0965\\
			\hline
			2 &0.8521	&0.8592	&0.9798	&0.8662	&0.8889	&0.9946	&1.0875	&1.0430\\
			\hline
			3 &0.8783	&0.9216	&1.0205	&0.9005	&0.8516	&0.9485	&1.0448	&1.0074\\
			\hline
			4 &1.0137	&1.0063	&1.0463	&0.9176	&0.8417	&0.9317	&1.0466	&1.0028\\
			\hline
			5 &1.0434	&1.0392	&1.0449	&0.9080	&0.8517	&0.9362	&1.0446	&1.0003\\
			\hline
		\end{tabular}}
		\label{table-CLSciSumm2017-SemanticSimilarity}
		\quad
	\end{subtable}
	\label{table-CLSciSumm2017}
\end{table}

\subsubsection{Results for CL-SciSumm 2017 dataset}
\label{subsubsec:perf-evaluation-CLSciSumm2017}
Tables \ref{table-CLSciSumm2017-Jaccard} - \ref{table-CLSciSumm2017-SemanticSimilarity} present the result of performance evaluation of \textit{P-Summ} on CL-SciSumm 2017 test set. It is evident from the table that both Jaccard Index (Table \ref{table-CLSciSumm2017-Jaccard}) and Jensen Shannon Distance (Table \ref{table-CLSciSumm2017-JSD}) exhibit trends similar to those exhibited by CL-SciSumm 2016 test set. We observe deviation from the ideal behaviour in the values for $\mathcal R^-$ and $\mathcal R^+$ for smaller summary sizes ($10\% \mbox{ and } 15\%$). This is attributed to the fact that for some signal strengths, it is either difficult to hold back the information not desired in the summary (for $\mathcal R^-$) or the generic summary of the document already contains the information desired by the user in the summary (for $\mathcal R^+$). Further, smaller length of the articles in CL-SciSumm 2017 dataset (Table \ref{table-datasets}) impedes inclusion of the information not desired by the user in the personal summary. 

\begin{figure}[h]
\centering
\begin{subfigure}{0.48\textwidth}
\centering
\includegraphics[width=\textwidth]{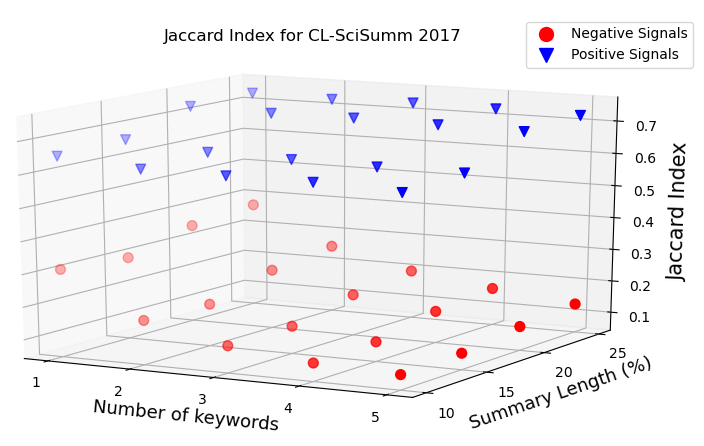}
\caption{Jaccard Index}
\label{fig-scatter-scisumm17-jaccard}
\end{subfigure}
\begin{subfigure}{0.48\textwidth}
\centering
\includegraphics[width=\textwidth]{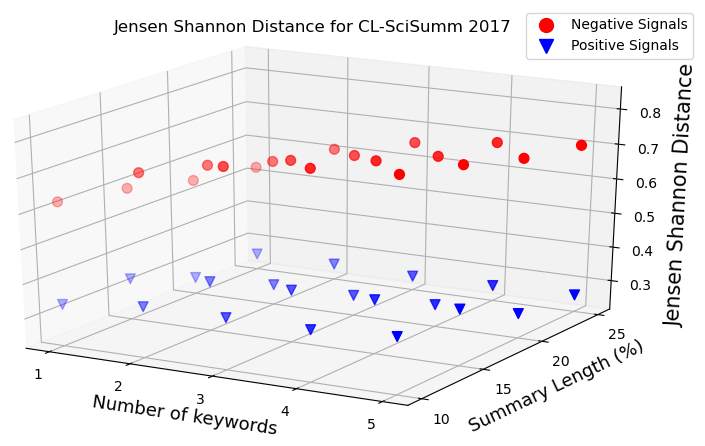}
\caption{Jensen Shannon Distance}
\label{fig-scatter-scisumm17-jsd}
\end{subfigure}
\caption{Comparison of performance for negative and positive knowledge signals in CL-SciSumm 2017 dataset based on Jaccard Index and Jensen Shannon Distance metrics}
\label{fig-scatter-scisumm17}
\end{figure}

Scatter plots in Fig. \ref{fig-scatter-scisumm17-jaccard} and \ref{fig-scatter-scisumm17-jsd} present the comparative analysis of \textit{P-Summ} for positive and negative knowledge signals on CL-SciSumm 2017 test set documents using Jaccard Index and Jensen Shannon distance. The contrast between the scores for positive and negative signals reveals that \textit{P-Summ} definitely eliminates the known knowledge and adds the desired knowledge in the personal summary.

\subsubsection{Results for CL-SciSumm 2018 dataset}
\label{subsubsec:perf-evaluation-CLSciSumm2018}
The results of CL-SciSumm 2018 test set for negative knowledge signals are presented in the first four columns of Tables \ref{table-CLSciSumm2018-Jaccard} - \ref{table-CLSciSumm2018-SemanticSimilarity}. We observe patterns that resemble the findings for CL-SciSumm 2016 and 2017 datasets.  Table \ref{table-CLSciSumm2018-Jaccard} shows that the values for Jaccard Index decrease with increase in signal strength, while Jensen Shannon Distance increases (Table \ref{table-CLSciSumm2018-JSD}). As expected, ratio $\mathcal R^-$ is less than one in all cases, demonstrating that negative knowledge signals are weaker in personal summary. These results affirm the competence of \textit{P-Summ} to meet the knowledge needs of the user by eliminating negative knowledge.

\begin{table}[ht]
	\centering
            \caption{Performance of \textit{P-Summ} on CL-SciSumm 2018 dataset. kw: number of keywords in $\mathcal K^-$ or in $\mathcal K^+$, $\mathcal D$: document length, JI: Jaccard Index, JSD: Jensen-Shannon distance, $\mathcal R$: ratio of semantic similarities (Eq. \ref{eqn-ratio-summ-knowledge})}
        \scriptsize
	\begin{subtable}[t]{\linewidth}
		\centering
        \caption{CL-SciSumm 2018 Jaccard Index}
		\resizebox{\textwidth}{!}{\begin{tabular}{c|c|c|c|c||c|c|c|c}
			\hline
			&\multicolumn{4}{c||}{Negative Signals}&\multicolumn{4}{c}{Positive Signals}\\
			\hline
			kw&{$10\%$ * $\mathcal{D}$}&{$15\%$ * $\mathcal{D}$}&{$20\%$ * $\mathcal{D}$}&{$25\%$ * $\mathcal{D}$}	&{$10\%$ * $\mathcal{D}$}&{$15\%$ * $\mathcal{D}$}&{$20\%$ * $\mathcal{D}$}&{$25\%$ * $\mathcal{D}$}\\
			\hline
			1 &0.3162	&0.3530	&0.3595	&0.3739	&0.6186	&0.6492	&0.7063	&0.7432\\
			\hline
			2 &0.2353	&0.2505	&0.2575	&0.2631	&0.6303	&0.6632	&0.7102	&0.7424\\
			\hline
			3 &0.2029	&0.2030	&0.2060	&0.2070	&0.6437	&0.6672	&0.7117	&0.7425\\
			\hline
			4 &0.1857	&0.1900	&0.1786	&0.1818	&0.6448	&0.6675	&0.7100	&0.7427\\
			\hline
			5 &0.1821	&0.1936	&0.1731	&0.1665	&0.6450	&0.6670	&0.7096	&0.7443\\
			\hline		
		\end{tabular}}
		\label{table-CLSciSumm2018-Jaccard}
		\quad
	\end{subtable}
	\\
        \vspace{2mm}
	\begin{subtable}[t]{\linewidth}
		\centering
        \caption{CL-SciSumm 2018 Jensen-Shannon Distance}
		\resizebox{\textwidth}{!}{\begin{tabular}{c|c|c|c|c||c|c|c|c}
			\hline
			&\multicolumn{4}{c||}{Negative Signals}&\multicolumn{4}{c}{Positive Signals}\\
			\hline
			kw&{$10\%$ * $\mathcal{D}$}&{$15\%$ * $\mathcal{D}$}&{$20\%$ * $\mathcal{D}$}&{$25\%$ * $\mathcal{D}$}	&{$10\%$ * $\mathcal{D}$}&{$15\%$ * $\mathcal{D}$}&{$20\%$ * $\mathcal{D}$}&{$25\%$ * $\mathcal{D}$}\\
			\hline
			1 &0.5877	&0.5373	&0.5287	&0.5038	&0.3403	&0.3478	&0.3053	&0.3043\\
			\hline
			2 &0.6550	&0.6244	&0.6107	&0.5865	&0.3280	&0.3419	&0.3069	&0.3243\\
			\hline
			3 &0.6882	&0.6720	&0.6536	&0.6330	&0.3127	&0.3430	&0.3072	&0.3568\\
			\hline
			4 &0.7034	&0.6888	&0.6763	&0.6573	&0.3097	&0.3448	&0.3107	&0.3765\\
			\hline
			5 &0.7067	&0.6903	&0.6859	&0.6709	&0.3075	&0.3432	&0.3142	&0.3895\\
			\hline
		\end{tabular}}
		\label{table-CLSciSumm2018-JSD}
		\quad
	\end{subtable}
	\\
        \vspace{2mm}
	\begin{subtable}[t]{\linewidth}
		\centering
        \caption{CL-SciSumm 2018 Ratio $\mathcal{R}$}
		\resizebox{\textwidth}{!}{\begin{tabular}{c|c|c|c|c||c|c|c|c}
			\hline
			&\multicolumn{4}{c||}{Negative Signals ($\mathcal R^-$)}&\multicolumn{4}{c}{Positive Signals ($\mathcal R^+$)}\\
			\hline
			kw&{$10\%$ * $\mathcal{D}$}&{$15\%$ * $\mathcal{D}$}&{$20\%$ * $\mathcal{D}$}&{$25\%$ * $\mathcal{D}$}	&{$10\%$ * $\mathcal{D}$}&{$15\%$ * $\mathcal{D}$}&{$20\%$ * $\mathcal{D}$}&{$25\%$ * $\mathcal{D}$}\\
			\hline
			1 &0.8381	&0.8347	&0.8540	&0.8827	&0.9021	&0.9515	&1.0050	&1.0208\\
			\hline
			2 &0.9622	&0.8941	&0.8884	&0.8972	&0.9474	&0.9401	&0.9701	&0.9795\\
			\hline
			3 &0.9020	&0.8731	&0.8757	&0.8722	&0.9785	&0.9958	&1.0202	&1.0188\\
			\hline
			4 &0.9275	&0.8697	&0.8673	&0.8694	&0.9911	&1.0130	&1.0341	&1.0250\\
			\hline
			5 &0.9641	&0.8914	&0.8806	&0.8778	&0.9981	&1.0264	&1.0384	&1.0291\\
			\hline		
		\end{tabular}}
		\label{table-CLSciSumm2018-SemanticSimilarity}
		\quad
	\end{subtable}
	\label{table-CLscisumm2018}
\end{table}

\begin{figure}[h]
\centering
\begin{subfigure}{0.48\textwidth}
\centering
\includegraphics[width=\textwidth]{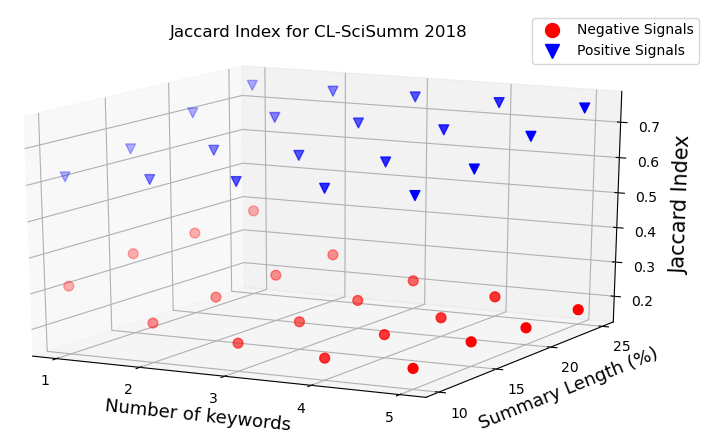}
\caption{Jaccard Index}
\label{fig-scatter-scisumm18-jaccard}
\end{subfigure}
\begin{subfigure}{0.48\textwidth}
\centering
\includegraphics[width=\textwidth]{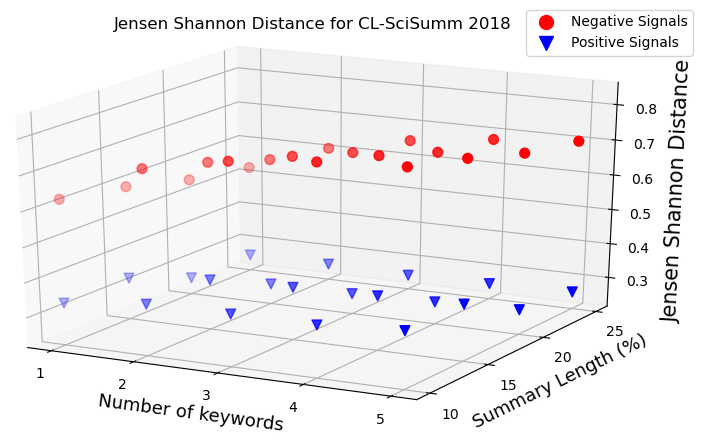}
\caption{Jensen Shannon Distance}
\label{fig-scatter-scisumm18-jsd}
\end{subfigure}
\caption{Comparison of performance for negative and positive knowledge signals in CL-SciSumm 2018 dataset based on Jaccard Index and Jensen Shannon Distance metrics}
\label{fig-scatter-scisumm18}
\end{figure}
The right four columns in Tables \ref{table-CLSciSumm2018-Jaccard} - \ref{table-CLSciSumm2018-SemanticSimilarity} show the results on CL-SciSumm 2018 test set for positive knowledge signals. We observe trends similar to CL-SciSumm 2016 and 2017 datasets for different signal strengths and summary lengths. Scatter plots in Fig. \ref{fig-scatter-scisumm18-jaccard} and \ref{fig-scatter-scisumm18-jsd} for comparative performance evaluation of \textit{P-Summ} algorithm on CL-SciSumm 2018 test set show the trend similar to CL-SciSumm 2016 and CL-SciSumm 2017 test sets. 

The results show that the expected trends are much clearer for AIPubSumm dataset than for CL-SciSumm datasets. This is possibly due to the fact that the author-defined keywords better represent the important topics discussed in the article than the extracted keywords. Algorithmically extracted keywords, on the other hand, are biased towards frequently occurring terms and phrases, and may not accurately reflect the main theme of an article. This also explains the notable divergence in the values of all three metrics for $\mathcal K^-$ and $\mathcal K^+$ for AIPubSumm dataset.  Hence, it is reasonable to conclude that author given keywords are of better quality than those extracted by algorithms in CL-SciSumm datasets. 

\textit{We have added the Personal Summary of the complete manuscript in the Appendix \ref{app:personal-summaries}, along with the computed metrics. The keyphrase  ``personal summary'' is  used as knowledge signal, and the summary length is $10\%$ of the original document.
}
\section{Conclusion and Future Directions}
\label{sec:conclusion}
\new{In this paper, we propose an unsupervised algorithm for extractive \textit{personalized} summarization task, which considers user preferences and extracts compatible content from the document to generate the user-specific \textit{personal} summary. The user expresses knowledge needs as \textit{positive} or \textit{negative} signals, which respectively correspond to the information \textit{desired} or \textit{not desired} in the summary. We represent the document as a binary term-sentence matrix and uncover the latent semantic space of the document using Weighted Non-negative Matrix Factorization. The weight matrix is judiciously set while taking into account the global and local context of the user-specified knowledge signals. The algorithm exploits the latent space to prioritize sentences in the personal summary by diminishing the score of sentences conveying knowledge \textit{not desired}, and boosting the score for those conveying \textit{desired} knowledge.}

\new{We also propose an evaluation framework, which uses system-generated \textit{generic} summary as the gold standard to assess the extent to which the user knowledge preferences have been accommodated in the algorithmic \textit{personal} summary. The proposed multi-granular framework evaluates the personal summary quality at sentence, term, and semantic level and is used for comprehensive evaluation of the \textit{P-Summ} algorithm for different summary lengths and signal strengths. Additionally, the proposed multi-granular evaluation framework averts the need for creating mode-specific datasets with mode-specific gold summaries for evaluation. The framework relies on a high-quality system-generated generic extractive summary, which faithfully conveys the important ideas of the document. We carry out an empirical investigation using scientific scholarly articles from four public datasets. The results demonstrate that the proposed algorithm effectively meets the knowledge needs of the user.} \\

   \noindent \new{\textbf{Future Directions:}  In the future, algorithmic improvements and model architectures beyond wNMF  need to be explored to further refine the latent semantic representation of terms and sentences. This may lead to enhanced effectiveness of \textit{personalized} summarization  algorithm. }

\new{The evaluation of a personal summary that integrates both negative and positive knowledge signals presents a significant challenge due to the inherent complexity of reconciling conflicting instructions. This necessitates the development of a novel and robust evaluation protocol capable of discerning the information added in response to positive knowledge signals and the information removed corresponding to negative knowledge signals. Such a protocol should consider factors like relevance, coherence, and user preferences, ensuring that the resultant personal summary strikes a balance between inclusivity and exclusivity, thereby meeting the user's knowledge requirements.}

\new{The diversity in the modes for expressing user knowledge preference semantic signals is an impediment in the growth of personal summarization research.  The lack of standard representation for different types of user knowledge signals complicates the comparison of personalized summarization algorithms. A plausible solution that can be explored in the future is to project the user preferences into the semantic space that can be understood by different summarization algorithms. The common representational mode for different types of knowledge signals eliminates the necessity for creating datasets tightly linked to the type of knowledge signals. This approach would not only streamline algorithm evaluation but also encourage researchers to focus on refining the summarization process.}

\section{Acknowledgements}
The authors gratefully acknowledge the time and effort by the anonymous reviewers. The probing questions, insightful comments, and suggestions by the reviewers have resulted in significant improvement of the paper.

\bibliographystyle{model5-names}
\biboptions{authoryear}
\bibliography{main}

\newpage

\begin{appendices}
\section{Appendix}
\label{sec-appendix}
\subsection{Semantic Similarity between Knowledge Signals and Personal Summary}
\label{appendix-semantic-similarity}
Let $K = \{k_1, k_2, \dots, k_s\}$ denote the set of $s$ uni-grams in user knowledge signals $\mathcal K$, and let $P = \{P_1, P_2,\dots, P_{l}\}$ represents the set of  $l$ sentences in the summary. Term-sentence matrix $A$ for document $D$ is decomposed into the product of matrices $U, V$ using standard NMF ($A \approx  UV$), which provides an unbiased latent representation of its terms and sentences.

The  $i^{th}$ row of matrix $U$, $(u_{i1}, u_{i2}, \dots, u_{ir})$, represents the latent semantic strength of term $k_i$ in $r$ latent topics\footnote{Approach to compute the number of latent topics is described in Sec. \ref{subsec:number-of-latent-factors}}. We obtain semantic representation of  $K$ by averaging the semantic strengths of all terms in $r$ latent topics and  denote it as $\hat K$. Thus $\hat K$ is row vector of $r$ elements, where the $i^{th}$ element is computed as $\frac{1}{s}\sum\limits_{p \in K} u_{pi}$.

Next, we compute semantic similarity of $\hat K$, with all sentences in the summary. Recall that columns in matrix $V$ represent the strength of sentences in $r$ latent topics. Since $\hat K$ and columns in $V$ are compatible vectors, it is legitimate to compute semantic similarity between knowledge signals and summary sentences. Averaging cosine similarities of $\hat K$ over $l$ sentences proxies for semantic similarity between $\mathcal K$ (represented as $\hat K$) and the summary ($S_g$ or $S_p$).
\begin{equation}\sigma(\mathcal K, Summary)  \approx \frac{1}{l}{ \sum\limits_{i=1}^{l} \mbox{cosine}(\hat K, P_{i})}\label{eqn-sim-sem}
\end{equation}
\subsection{Personal Summary of the Manuscript}
\label{app:personal-summaries}
	\begin{figure}[H]
 		\centering
 		\caption{Generic and \textit{P-Summ} algorithm summaries of the manuscript. Keyword `\textit{personal summary}' is given as user knowledge signal ($\mathcal K^+$ and $\mathcal K^-$). Extracted summary length is $10\%$ of document length.}
 		\begin{subfigure}[b]{\textwidth}
 			\caption{Generic Summary of the manuscript}
 			\label{app:fig-generic-summ-1}
 			\noindent\fbox{%
 				\parbox{\textwidth}{%
 					\scriptsize{\setstretch{1.0}{
{$S_1$}: The proposed algorithm, P-Summ,  permits the user to supply both positive and negative knowledge signals and creates the document summary that ensconces content semantically similar to the positive signals and dissimilar to the negative signals.\\
{$S_2$}: The task of  personalized summarization entails two sub-tasks, viz., (i) modeling the user's knowledge preferences to be communicated to the algorithm and  (ii) creating a desired summary using either extractive or abstractive (neural) algorithm to best match the communicated knowledge preferences. \\
{$S_3$}: To overcome the above-mentioned challenges, we propose  P-Summ algorithm that accepts keywords or phrases as positive or negative  knowledge signals to create a \textbf{personal} summary. \\
{$S_4$}:	We propose a Personalized Summarization algorithm, P-Summ, which is competent to handle both negative and positive knowledge signals specified by the user to create a \textbf{personal summary}.\\
{$S_5$}:	The algorithm elicits knowledge that is semantically related to the user specified signals in the local and global context of the document and leverages it to meet user expectations in the \textbf{personal summary}.\\
{$S_6$}:	The user knowledge signals and binary term sentence matrix of the document are used to construct the weight matrix, which plays an instrumental role in identifying information desired and not desired in the summary.\\
{$S_7$}:	We categorize the approaches for personalized summarization on the basis of mode of user knowledge signals and type of summarization algorithm adopted to generate \textbf{personal summary} of the document, and describe them in detail below.\\
{$S_8$}:	QFS methods generate \textbf{personal summary} of a document based on user knowledge signals communicated as natural language queries.\\
{$S_9$}:	An effective \textbf{personal summary} must not only avoid  the content directly conveyed by the user specified knowledge signals, but must also consider other semantically related content in the document.\\
{$S_{10}$}:	Post factorization, strengths of terms and sentences in the latent semantic topics are reduced  to favour their exclusion  in the \textbf{personal summary} based on the type of user knowledge signal.\\
{$S_{11}$}:	Since the role of the terms corresponding to user knowledge signal for generating \textbf{personal summary} is paramount, we follow the term based sentence scoring method proposed earlier by Khurana \& Bhatnagar (2019).\\
{$S_{12}$}:	We consider author defined keywords as prominent topics or concepts described in a scientific article and use them to simulate user knowledge signals for generating \textbf{personal summaries}.\\
{$S_{13}$}:	For each combination of signal strength and summary length, persistently higher values of the Jaccard Index for $\mathcal K^+$ compared to the corresponding entries for $\mathcal K^-$ suggest that the \textbf{personal summary} corresponding to positive knowledge decisively ensconces the information desired by the user .\\
{$S_{14}$}:	This asserts that when the same set of keywords is used for simulating both types of knowledge signals, P-Summ algorithm assertively includes the terms desired by the user and omits the unwanted keywords and other related terms.\\
{$S_{15}$}: The clear separation of the data points for positive and negative knowledge signals reveals the efficacy of the proposed algorithm in generating user specific summaries.\\
{$S_{16}$}:	In this paper, we propose an unsupervised algorithm for extractive personalized summarization task, which considers user preferences and extracts compatible content from the document to generate the user specific \textbf{personal summary}.\\
{$S_{17}$}:	We also propose an evaluation framework, which uses system generated generic summary as the gold standard to assess the extent to which the user knowledge preferences have been accommodated in the algorithmic \textbf{personal summary}.\\
{$S_{18}$}:	The proposed multi-granular framework evaluates the \textbf{personal summary} quality at sentence, term, and semantic level and is used for comprehensive evaluation of the P-Summ algorithm for different summary lengths and signal strengths.	
}}}}
 		\end{subfigure}
\end{figure}
\begin{figure}\ContinuedFloat
 		\begin{subfigure}[p]{\textwidth}
 			\caption{\textit{P-Summ} summary when keyword `\textit{personal summary}' is given as positive knowledge signal, $\mathcal K^+$. Jaccard Index between the sentences of generic and \textit{P-Summ} algorithm summaries is $0.895$. Jensen Shannon Distance between distribution of terms in generic and \textit{P-Summ} algorithm summaries is $0.152$. The value of similarity ratio $\mathcal R^+ = 1.039 (> 1)$}
 			\label{app:fig-positive-summ-1}
 			\noindent\fbox{%
 				\parbox{\textwidth}{%
 					\scriptsize{\setstretch{1.0}{
     {$S_1$}: The proposed algorithm, P-Summ,  permits the user to supply both positive and negative knowledge signals and creates the document summary that ensconces content semantically similar to the positive signals and dissimilar to the negative signals.\\
{$S_2$}: The task of  personalized summarization entails two sub-tasks, viz., (i) modeling the user's knowledge preferences to be communicated to the algorithm and  (ii) creating a desired summary using either extractive or abstractive (neural) algorithm to best match the communicated knowledge preferences. \\
{$S_3$}: To overcome the above-mentioned challenges, we propose  P-Summ algorithm that accepts keywords or phrases as positive or negative  knowledge signals to create a \textbf{personal summary}. \\
{$S_4$}: We propose a Personalized Summarization algorithm, P-Summ, which is competent to handle both negative and positive knowledge signals specified by the user to create a \textbf{personal summary}.\\
{$S_5$}:	The algorithm elicits knowledge that is semantically related to the user specified signals in the local and global context of the document and leverages it to meet user expectations in the \textbf{personal summary}.\\
{$S_6$}:	The user knowledge signals and binary term sentence matrix of the document are used to construct the weight matrix, which plays an instrumental role in identifying information desired and not desired in the summary.\\
{$S_7$}:	We categorize the approaches for personalized summarization on the basis of mode of user knowledge signals and type of summarization algorithm adopted to generate \textbf{personal summary} of the document, and describe them in detail below.\\
{$S_8$}:	QFS methods generate \textbf{personal summary} of a document based on user knowledge signals communicated as natural language queries.\\
{$S_9$}:	A \textbf{personal summary} of a document aims to capture the vital content from the document to meet the topical knowledge needs of the user, which may be expressed as knowledge signals .\\
{$S_{10}$}:	An effective \textbf{personal summary} must not only avoid  the content directly conveyed by the user specified knowledge signals, but must also consider other semantically related content in the document.\\
{$S_{11}$}:	Post factorization, strengths of terms and sentences in the latent semantic topics are reduced  to favour their exclusion  in the \textbf{personal summary} based on the type of user knowledge signal.\\
{$S_{12}$}:	Since the role of the terms corresponding to user knowledge signal for generating \textbf{personal summary} is paramount, we follow the term based sentence scoring method proposed earlier by Khurana \& Bhatnagar (2019).\\
{$S_{13}$}:	We consider author defined keywords as prominent topics or concepts described in a scientific article and use them to simulate user knowledge signals for generating \textbf{personal summaries}.\\
{$S_{14}$}:	For each combination of signal strength and summary length, persistently higher values of the Jaccard Index for $\mathcal K^+$ compared to the corresponding entries for $\mathcal K^-$  suggest that the \textbf{personal summary} corresponding to positive knowledge decisively ensconces the information desired by the user .\\
{$S_{15}$}:	The clear separation of the data points for positive and negative knowledge signals reveals the efficacy of the proposed algorithm in generating user specific summaries.\\
{$S_{16}$}:	In this paper, we propose an unsupervised algorithm for extractive personalized summarization task, which considers user preferences and extracts compatible content from the document to generate the user specific \textbf{personal summary}.\\
{$S_{17}$}:	We also propose an evaluation framework, which uses system generated generic summary as the gold standard to assess the extent to which the user knowledge preferences have been accommodated in the algorithmic \textbf{personal summary}.\\
$S_{18}$:	The proposed multi-granular framework evaluates the \textbf{personal summary} quality at sentence, term, and semantic level and is used for comprehensive evaluation of the P-Summ algorithm for different summary lengths and signal strengths.
}}}}
 		\end{subfigure}
 		\end{figure}
   \begin{figure}\ContinuedFloat
 		\begin{subfigure}[p]{\textwidth}
 			\caption{\textit{P-Summ} summary when keyword `\textit{personal summary}' is given as negative knowledge signal, $\mathcal K^-$. Jaccard Index between the sentences of generic and \textit{P-Summ} algorithm summaries is $0.0$. Jensen Shannon Distance between distribution of terms in generic and \textit{P-Summ} algorithm summaries is $0.780$. The value of similarity ratio $\mathcal R^- = 0.631 (< 1)$}
 			\label{app:fig-negative-summ-1}
 			\noindent\fbox{%
 				\parbox{\textwidth}{%
 					\scriptsize{\setstretch{1.0}{{$S_1$}: Using this framework, we conduct a comprehensive evaluation of P-Summ algorithm on four publicly available datasets consisting of scientific scholarly articles.\\
{$S_2$}:	We also propose a multi-granular evaluation framework and demonstrate the effectiveness of P-Summ algorithm for personal summarization of scientific scholarly articles, which are often lengthy and contain multiple points of interest.\\
{$S_3$}:	We use Weighted Non-negative Matrix Factorization  to expose the latent semantic space of the document and process both positive and negative knowledge signals in this space .\\
{$S_4$}:	We state the experimental design in Sec 6, and present the results of comprehensive evaluation of P-Summ algorithm on four public datasets comprising scientific scholarly articles in Sec 7.\\
{$S_5$}:	We choose to go with relatively recent method of Weighted Non-negative Matrix Factorization  because of its ability to divulge the importance of terms and sentences in the latent semantic space of the document in accordance with the weight matrix.\\
{$S_6$}:	Let $W_{m \times n}$ be a matrix of non negative weights, where the entry $w_{ij}$ controls the prominence of the element $a_{ij}$ in the latent space.\\
{$S_7$}:	In the current context, matrix $A$ is the binary incidence matrix of the document, matrix $U$ captures relationship among the terms and latent factors, and matrix $V$ represents relationship among the latent factors and sentences in the latent semantic space.\\
{$S_8$}:	In the current context, eigen centrality of the nodes in $\mathcal G$ quantifies importance of terms in the global context of the document.\\
{$S_9$}:	Given the term $t_i \in \mathcal K$ and the corresponding node $n_i \in \mathcal G$, we denote its eigen centrality by $\epsilon_i$ signifying the global importance of $t_i$.\\
{$S_{10}$}:	Conjecturing that the penalization  of the terms should be in conformity with their importance in both local and global context of the document text, we propose the following weighting scheme.\\
{$S_{11}$}:	Step 2 computes the eigen centrality of nodes in graph.\\
{$S_{12}$}:	In step 5, the algorithm creates the induced sub-graph $\mathcal H$ for terms in cohort $\mathbb K$, and step 6 computes the degree of nodes.\\
{$S_{13}$}:	Since construction of the weight matrix takes into account the local and global context of the user given knowledge signals, we expect the new content to be in consonance with the given knowledge signals.\\
{$S_{14}$}:	We present the results for each dataset in three tables corresponding to the three metrics viz Jaccard Index, Jensen Shannon Distance and ratio R, described in Sec. \ref{sec:evaluation-framework}.\\
{$S_{15}$}:	Tables 3a-3c present the results for personal summaries for the articles in AIPubSumm test set, where author defined keywords given with the articles serve as knowledge signals.\\
{$S_{16}$}:	Plots in Figures 6a and 6b present comparative performance of P-Summ for positive and negative knowledge signals on CL-SciSumm 2016 documents using Jaccard Index and Jensen Shannon Distance, respectively.\\
{$S_{17}$}:	It is evident from the table that both Jaccard Index  and Jensen Shannon Distance  exhibit trends similar to those exhibited by CL-SciSumm 2016 test set.\\
{$S_{18}$}:	Scatter plots in Fig. 7a and 7b presents the comparative analysis of P-Summ for positive and negative knowledge signals on CL-SciSumm 2017 test set documents using Jaccard Index and Jensen Shannon distance.\\
{$S_{19}$}:	Table 6a shows that the values for Jaccard Index decreases with increase in signal strength, while Jensen Shannon Distance increases (Table \ref{table-CLSciSumm2018-JSD}).\\
{$S_{20}$}:	We represent the document as a binary term sentence matrix and uncover the latent semantic space of the document using Weighted Non-negative Matrix Factorization.
}}}}
 		\end{subfigure}
 		\label{app:fig-psumm-summaries-1}
 	\end{figure}

\end{appendices}
\end{document}